%% file: main.tex
\RequirePackage{fix-cm}
\documentclass[smallextended,natbib]{svjour3}       
\smartqed  
\usepackage{natbib}
\usepackage{graphicx}
\usepackage{url}
\usepackage{tcolorbox}
\usepackage{multirow}
\usepackage{amssymb,amsfonts}
\usepackage{pifont}
\usepackage{xcolor,colortbl} 
\usepackage{listings}
\usepackage{hyperref}
\usepackage{amsmath,amsfonts}
\usepackage{pdflscape}
\usepackage{graphicx}
\usepackage{soul}
\usepackage{pdflscape}
\usepackage{subcaption}
\usepackage{multirow}
\usepackage{adjustbox}
\usepackage{comment}
\usepackage{wrapfig}
\sethlcolor{black}
\usepackage{booktabs}
\usepackage{rotating}
\usepackage{siunitx}
\usepackage{breakurl} 
\usepackage{comment}
\usepackage{listings}
\usepackage{threeparttable}

\usepackage{placeins} 
\input{tex/commands}

\begin{document}


\title{Exploring User Privacy Awareness on \GH: An Empirical Study}


\titlerunning{Exploring User Privacy Awareness on \GH: An Empirical Study}        


\author{Costanza Alfieri \and Juri {Di Rocco} \and Paola Inverardi \and Phuong T. Nguyen}

\institute{Costanza Alfieri \at
		DISIM - University of L'Aquila (Italy)\\
	\email{costanza.alfieri@student.univaq.it} \and
        Juri Di Rocco, Phuong T. Nguyen \at
		DISIM - University of L'Aquila (Italy)\\
	\email{juri.dirocco@univaq.it}, \email{phuong.nguyen@univaq.it} \and
        Paola Inverardi \\
        Gran Sasso Science Institute (Italy)\\
	\email{paola.inverardi@gssi.it}
}

\maketitle

\begin{abstract}
	
    

 \GH provides developers with a practical way to distribute source code and collaboratively work on common projects. 
To enhance account security and privacy, \GH allows its users to manage access permissions, review audit logs, and enable two-factor authentication. However, despite the endless effort, 
the platform still faces various issues related to the privacy of its users. 
  This paper presents an empirical study delving into the GitHub ecosystem. Our focus is on investigating the utilization of privacy settings on the platform and identifying various types of sensitive information disclosed by users. 
  Leveraging a dataset comprising \numDevs developers, we report and analyze their activities by means of comments on pull requests. Our findings indicate an active engagement by users with the available privacy settings on GitHub. Notably, we observe the disclosure of different forms of private information within pull request comments. This observation has prompted our exploration into sensitivity detection using a large language model and BERT, to pave the way for a personalized privacy assistant. \revised{Our work provides insights into the utilization of existing privacy protection tools, such as privacy settings, along with their inherent limitations. Essentially, we aim to advance research in this field by providing both the motivation for creating such privacy protection tools and a proposed methodology for personalizing them.} 


\end{abstract}


\section{Introduction}
\label{sec:Introduction}
\input{tex/intro_new}

    \section{Motivations and Background}
\label{sec:MotivatingExample}
\input{tex/MotivatingExample}

\section{Related work}
\label{sec:RelateWork}
\input{tex/RelatedWork}

\section{Methodology}
\label{sec:Methodology}
\input{tex/methodology}

\section{Sensitivity detection in textual comments}
\label{sec:prediction}

\input{tex/prediction}

\section{Empirical results}
\label{sec:EmpiricalResults}
\input{tex/empirical_results}

\section{Discussion}
\label{sec:Discussion}

\input{tex/discussion}

\section{Conclusion \revisedtwo{and Future Work}}
\label{sec:Conclusion}
\input{tex/conclusion}

\section*{Conflict of Interests}
\label{sec:statement}
\input{tex/statement}

\section*{Data Availability Statements }
\label{sec:das}
\input{tex/das}
	
	\begin{acknowledgements}
This work has been partially supported by: MUR project 2020 
\emph{``EMELIOT"} grant n. 2020W3A5FY; MUR projects PRIN 2022 PNRR: \emph{``FRINGE: context-aware FaiRness engineerING in complex software systEms''} grant n. P2022553SL, \emph{``TRex-SE: Trustworthy Recommenders for Software Engineers''} grant n. 2022LKJWHC, and \emph{``HALO: etHical-aware AdjustabLe autOnomous systems"} grant n. 2022JKA4SL. We also acknowledge the MUR Department of Excellence 2023 - 2027 program.

We thank the anonymous reviewers for their useful comments and suggestions that helped us improve our manuscript.
	\end{acknowledgements}

	%
	%

        \bibliographystyle{spbasic} 

\bibliography{tex/main}
\end{document}

%% file: tex/commands.tex
\usepackage{xspace}
\usepackage[colorinlistoftodos]{todonotes}

\newcommand*{\ie}{i.e.,\@\xspace}
\newcommand*{\eg}{e.g.,\@\xspace}

\newcommand*{\bertft}{\texttt{BERT$_{ft}$}\@\xspace}
\newcommand*{\llamaft}{\texttt{Llama$_{ft}$}\@\xspace}
\newcommand*{\llamazs}{\texttt{Llama$_{zs}$}\@\xspace}

\newcommand*{\GH}{GitHub\@\xspace}

\newcommand*{\CO}{\texttt{commits}\@\xspace}
\newcommand*{\CC}{\texttt{commit\_comments}\@\xspace}
\newcommand*{\FO}{\texttt{followers}\@\xspace}
\newcommand*{\PRC}{pull\_requests comments\@\xspace}
\newcommand*{\IC}{\texttt{issue\_comments}\@\xspace}

\newcommand*{\numDevs}{6,132\@\xspace}

\newcommand{\rqfirst}{\textbf{RQ$_1$}: \emph{Are the privacy settings provided on \GH used/adopted by the users? In other terms, do we observe different combinations of these settings, or is there a dominant configuration?} } 

\newcommand{\rqsecond}{\textbf{RQ$_2$}: \emph{What types of private information are disclosed on GitHub by users?}}

\newcommand{\rqthird}{\textbf{RQ$_3$}: \emph{After users choose their privacy settings, do they adhere to what they have declared? In other words, can we observe a discrepancy between their stated privacy preferences and their actual behavior, such as their textual activity?}}

\newcommand{\rqfourth}{\textbf{RQ$_4$}: \emph{
To which extent is it possible to automate the detection of sensitive comments with the use of BERT or \revisedtwo{Llama2?}}}

\makeatletter
\newcommand*{\etc}{%
	\@ifnextchar{.}%
	{etc}%
	{etc.\@\xspace}%
}
\makeatother

\newcommand\revised[1]{#1}
\newcommand\revisedtwo[1]{#1}

\newcommand{\code}[1]{{\texttt{#1}}}

\definecolor{darkgray}{gray}{0.78}
\definecolor{lightgray}{gray}{0.85}
\definecolor{verylightgray}{gray}{0.95}
\definecolor{codegreen}{rgb}{0,0.6,0}
\definecolor{codegray}{rgb}{0.5,0.5,0.5}
\definecolor{codepurple}{rgb}{0.58,0,0.82}
\definecolor{backcolour}{rgb}{0.95,0.95,0.92}
\lstdefinestyle{java}{
	backgroundcolor=\color{backcolour},   commentstyle=\color{codegreen},
	keywordstyle=\color{magenta},
	numberstyle=\tiny\color{codegray},
	stringstyle=\color{codepurple},
	basicstyle=\ttfamily\scriptsize,
	breakatwhitespace=false,         
	breaklines=true,                 
	captionpos=b,                    
	keepspaces=true,                 
	numbers=left,                    
	numbersep=5pt,                  
	showspaces=false,                
	showstringspaces=false,
	showtabs=false,                  
	tabsize=2
}

%% file: tex/intro_new.tex
In recent years, privacy in the digital world has become a major concern. Regulations like the EU General Data Protection Regulation (GDPR), the California Consumer Privacy Act (CCPA), or the UK's Data Protection Act have been ratified to 
regulate the (ab-)use of sensitive data~\citep{voigt2017eu,pardau2018california,doi:10.1080/713673366}. While being undeniably valuable, it has become clear that regulations alone may not suffice to ensure robust protection of user privacy. 
Indeed, software platforms provide mechanisms that allow users to set their privacy preferences and that, together with regulation should offer a wider shield to user privacy.

GitHub\footnote{\url{https://github.com/}} is a  platform for software developers where a significant amount of professional collaboration and coding takes place. As a matter of fact, it holds vast amounts of data that can be privacy sensitive for its users.
 GitHub users are first class inhabitants of the digital world and may be considered experts. However, ensuring that they are aware of the sensitivity of personal information that they leave on the platform and that they have control over who can see this data and how it's used, is crucial for maintaining trust in the platform and protecting users’ interests and safety. Indeed, GitHub offers to users means to declare which pieces of personal information they are willing to make public setting this way their privacy preferences. However, this is not enough provided that in their daily activity users may leave intentionally or unintentionally breadcrumbs of personal information open to the public. 

In this paper, we show that diverse pieces of personal information about a specific user can be uncovered on \GH, and this information may disclose data that are hidden in the user’s privacy settings. In the study, we extend upon concepts introduced in previous research. In particular, we take privacy settings provided on GitHub as a declaration of intent of the user that defines their privacy profile \citep{inverardi2023systematic, migliarini2020}. Our exploration focuses on determining whether these privacy settings are actively utilized by users and how. We leverage this information to compare it with the actual behaviors of users, particularly what they disclose in their comments. Our objective is to enhance our understanding of the effectiveness of privacy settings and \revised{explore any potential correlation with the users' observable behaviors. 
Our research offers insights into the utilization of existing privacy protection tools, such as privacy settings, along with their inherent limitations. Given the interest in developing privacy assistants (e.g., \cite{DBLP:conf/soups/0017ASAZSAA16}), our research supports this field by providing both the motivation for creating such privacy protection tools and a proposed methodology for personalizing them.}

To achieve this, we conducted an analysis of pull\_request comments within a subset of \GH users, namely the \textit{Active users} (see Section \ref{subsec:data_curation}), intending to identify what type of private information can be found on the platform. \revised{The choice of restricting to pull\_request is motivated by previous studies in \citep{sajadi2023interpersonal, iyer2019effects} (see Section \ref{sec:MotivatingExample})}. Our examination yielded insights into users' family info, moral values, workplace details, travel habits, time zones, and more. In conjunction with these findings, we present a dataset comprising both sensitive and non-sensitive comments, each manually labeled into distinct sensitive classes. 

In the analysis process, we curated a labeled dataset of pull\_requests comments made by the \textit{Active users.} The novelty of this labeled dataset consists in having labeled \emph{``digital behaviour'',} that can be linked to certain privacy profiles. 
The analysis discovered several discrepancies between the user privacy settings and pieces of information disclosed by the user in some comments. These discrepancies may represent instances of the so-called \emph{``privacy paradox,''} \ie inconsistencies between users declared preferences and their actual behavior~\citep{BARTH20171038,kokolakis2017privacy}, where users appear to be very attentive about their privacy, but eventually in their actual behavior 
miss to defend their personal data. Alternatively, they may simply indicate lack of attention/importance of the user for the privacy settings regarding her actual behavior. In both cases, the results show that the way personal privacy is dealt with in GitHub needs to be improved. 

To address similar problems, many researchers claim the importance of having automated tools to protect users' privacy in the digital world \citep{DBLP:conf/soups/0017ASAZSAA16,fukuyama2021save, DBLP:journals/access/AutiliRIPT19}. 

Our work represents a step forward in this direction allowing for the realization of an awareness tool that can assist the users in composing her textual comments in a privacy consistent way with the user profile. A proof of concepts of such a tool is presented in Section \ref{sec:prediction}.

To summarize, in this work we answer the following research questions \revised{that help understanding the privacy dynamics on \GH}: 

\begin{itemize}
    \item \revised{\rqfirst~} We investigate a large set of developers to find out if there exist differences in their privacy preferences. \revised{The use of privacy settings and the potential differences between users' selection indicate a certain attention to their privacy.}

    \item \rqsecond~Although designed as a platform for technical purposes, GitHub inherently possesses social media characteristics. Consequently, our study investigates the information that developers disclose, whether intentionally or unintentionally, in their pull\_requests comments.
 
    \item \revised{\rqthird~It is relevant to understand whether there is a mismatch to assess the effectiveness of these privacy settings and to facilitate the development of personalized privacy assistant for users.}
    \item \revised{\rqfourth }~\revised{We explored if the sensitivity detection of textual comments on this platform can be conducted by leveraging models as BERT or Llama2. }

\end{itemize}


The main contributions of our work are summarized below:
\begin{itemize}
    \item We conducted an empirical study on a set of \numDevs developers to investigate 
    \revised{the privacy dynamics} in the \GH ecosystem.
    \item \revised{Our investigation reveals that users adopt different configurations of privacy settings, showing different privacy concerns.}
    \item \revised{The empirical study shows that 
    users 
    reveal different types of information about themselves or other developers through pull\_request comments.
    This triggers the need to overhaul the privacy management from the designers of \GH.} 
    
    \item \revised{Our ultimate goal is to encourage the development of innovative applications for detecting privacy data leakage. Specifically, we present a proof of concept where we train BERT and we fine-tuned 
    Llama--a large language model \citep{touvron2023llama} (LLM)--to automatically assess whether a text discloses sensitive information. 
    }
    \item The dataset curated through this paper has been published\footnote{\url{https://github.com/MDEGroup/EMSE-CHASE-Privacy}} to facilitate future research (see Section \ref{sec:das}). 
\end{itemize}

The paper is structured as follows. In Section~\ref{sec:MotivatingExample}, we provide a background on privacy settings on \GH, a set of vulnerabilities that show the need for techniques and tools to manage privacy settings in the platform, \revised{and a real case scenario}.
Section~\ref{sec:RelateWork} reviews related work. 
Section~\ref{sec:Methodology} presents the methodology to perform the empirical study on the \GH ecosystem. Section~\ref{sec:prediction} presents the proof of concepts with BERT and Llama2. The results are reported and analyzed in Section~\ref{sec:EmpiricalResults}. We 
discuss the results in Section~\ref{sec:Discussion}. 
Section~\ref{sec:Conclusion} sketches future work, and concludes the paper.

%% file: tex/MotivatingExample.tex
In this section, we illustrate the different motivations behind this study and provide background on privacy settings on \GH and on \GH privacy vulnerabilities. 
\subsection{GitHub and its privacy settings} 
We chose to investigate GitHub for different reasons: 
\emph{(i)} \GH is a platform for technical purposes and may appear 
neutral or devoid of any privacy or ethical concerns. However, ``[open source software] is as much social as it is technical'' 
~\citep{vasilescu2015data} and there is already evidence from the literature that proves privacy infringements and gender discrimination on the platform~\citep{terrell2017gender,meliHowBadCan2019,fordCodeItselfHow2019,niuCodexLeaksPrivacyLeaks}; \emph{(ii)} On this platform, there is a large availability of data that can be downloaded as well as dataset already processed~\citep{gousios2013ghtorent}; \revised{ \emph{(iii)} Furthermore, this platform 
allows us to access users' privacy settings thus permitting to compare users' stated privacy preferences with their actual behaviors. This approach represents a novelty compared to previous studies on the utilization of privacy settings on platforms, which relied solely on user surveys rather than an analysis of actual preference selections~\citep{kanampiu2019privacy,chen2019demystifying}. The analysis of privacy preferences offers a deeper understanding of user behaviors and the limitations associated with the available privacy settings.}

In this study, \revised{
we concentrated on examining users' selection of privacy preferences and their self-disclosure of personal information in pull\_request comments. Our objective is to analyze the privacy dynamics on \GH, i.e., whether users effectively used the privacy settings provided, if these settings present any limitations, and if any personal information, intentionally concealed or not, can be inferred from a user's textual activity on \GH.}
Textual activities vary from making a commit, sending a pull\_request, commenting on all these different activities. 
In this study, we focused on the task of commenting on a pull\_request. This decision was guided by previous studies highlighting a greater probability of encountering significant user interactions and consequently more sensitive information in pull\_request contexts \citep{sajadi2023interpersonal, iyer2019effects}.

\begin{figure}[htp!]
    \centering
    \includegraphics[width=0.88\linewidth]{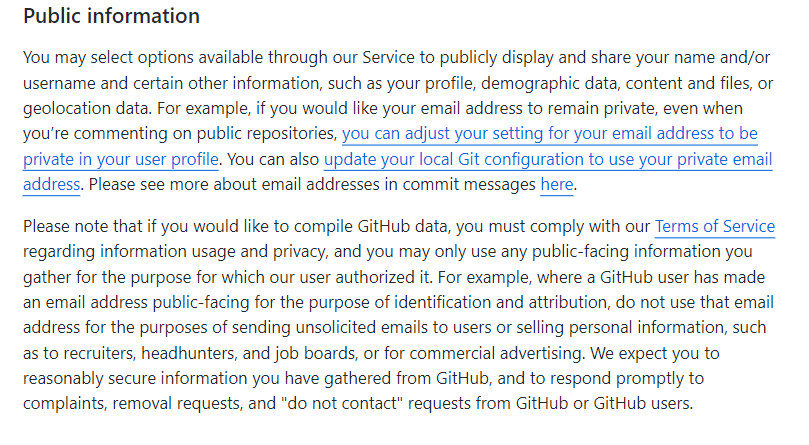}
    \caption{GitHub Privacy Statement on settings.}
    \label{fig:privacystatement}
\end{figure}

According to the \href{https://docs.github.com/en/site-policy/privacy-policies/github-privacy-statement#changes-to-our-privacy-statement}{GitHub Privacy Statement} (see also Figure \ref{fig:privacystatement}),\footnote{ \url{https://bit.ly/461tL9i}} a user who does not wish to show all the information available on her GitHub profile, can adjust the privacy settings provided by the platform in order to hide pieces of information.

\begin{figure}[!htp]
 \begin{subfigure}{0.4\textwidth}
    \centering
    \includegraphics[width=\linewidth]{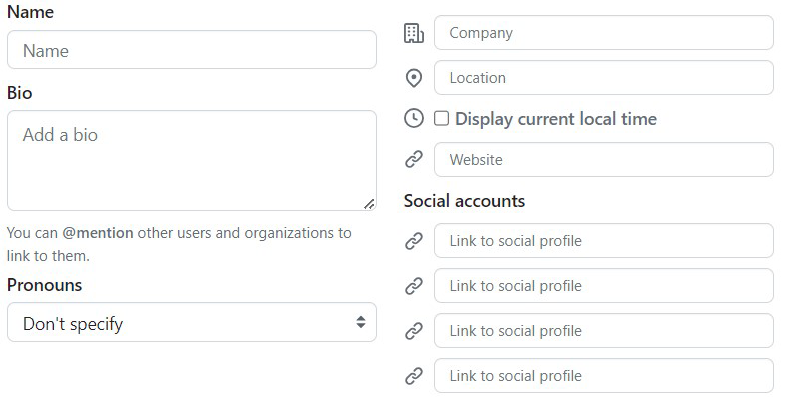}
    \caption{Personal information}
   \end{subfigure}
    \hfill
    \begin{subfigure}{0.6\textwidth}
        \centering
        \includegraphics[width=1\linewidth]{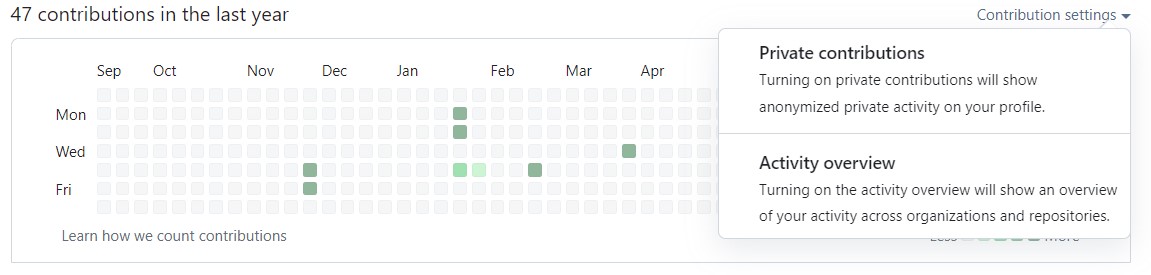}
        \caption{Contribution settings on GitHub}
        \label{fig:contribution}
    \end{subfigure}
    \caption{Privacy settings on GitHub.}\label{fig:privacy}
 \end{figure}

When referring to users' privacy, we talk about \textit{personal} users' privacy, \ie what they aim to share about their life. On GitHub, these \textit{desiderata} can be expressed through the privacy settings of the profile. An example of what can be shown or hidden on GitHub is illustrated in Figure \ref{fig:privacy}. We consider the totality of the privacy settings selected by the users as their \textit{privacy desiderata}. 

\subsection{Privacy vulnerability in \GH}

Privacy vulnerability in \GH arise in several places.
Personal information such as email addresses, location, and potentially even real names are often part of a user's profile. 
Even if \GH provides tailored settings for privacy (see Figure~\ref{fig:privacy}), 
the activities of users, such as the repositories they star, the issues they comment on, or the projects they follow, can reveal a lot about their interests, expertise, and professional activities. 
Due to the social nature of \GH, user collaborations can be used to identify user's professional network, which might include sensitive information, especially for those working on confidential or competitive projects.

Privacy vulnerabilities 
can arise 
as follows: 

\begin{itemize}
    \item [V1] The activities of a user who has chosen to set her event settings to private can potentially be discovered by examining the history of repositories to which she has made contributions.
    \item [V2] When users make links that expose personal or sensitive information, their \emph{privacy desiderata} may be compromised.
    \item [V3]  Public visibility occurs when users fork or star each other's repositories. These actions  
provide information about users' hobbies, projects, or collaborations. Additionally, issue and pull request discussions are public by default, 
revealing complex collaboration networks. 
A user frequently addressing user-initiated issues or pull requests,  may suggest a relationship beyond \GH, such as coworkers or acquaintances. Many organizations and teams are open to the public. This could reveal confidential professional relationships or collaborations. The public exposure of the user's contribution graph, which shows their work patterns and hours, might reveal work habits and inactivity. This information may be linked to job absence or personal activities.
    \item [V4] 
Textual contents such as commit messages, pull request, and issue comments may disclose privacy information.

\end{itemize}

The above listed privacy vulnerabilities are well known to researcher who publish under double-blind peer review contexts~\citep{7965366}.
In 
many research environments, double-blind peer review is pivotal to the integrity of knowledge dissemination. This requires to ensure the anonymity of submission materials, including associated replication packages. 
On GitHub, users need to anonymize sensitive information to protect privacy while still allowing for the replication of their results. 
Although \GH allows users to specify privacy settings for their repositories reducing the likelihood of disclosing user identities, it is necessary to thoroughly analyze repository content, metadata, and textual comments (\eg commit, issue, and pull request comments) to delete any trace that can reveal identities or affiliations. 
In this context, external tools are developed to remove user data from repositories, \eg \emph{Anonymous GitHub}.\footnote{\url{https://anonymous.4open.science/}} However, the risk of revealing the user identity still exists as those tools mainly apply rewriting rules, where the user should provide the complete list of terms that will be anonymized. Moreover, to the best of our knowledge, no tool is provided for sanitizing textual comments which represent a threat to privacy, as we show in this paper. 
Ideally, any automated means should take in input the user's privacy desiderata and then act to support the user in maintaining her privacy desiderata throughout her interactions with the system.

\subsection{The recruiters problem}
It is widely acknowledged that recruiters leverage social networks to gather insights into their candidates' backgrounds \citep{bectonSocialMediaSnooping2019,elouirdiRelationshipRecruiterCharacteristics2016}. The information sought varies, including personality traits, communication skills, the presence of provocative or inappropriate photographs, or any other factors that might dissuade the hiring of a candidate\citep{hendersonTheyPostedWhat2019}.


A survey conducted by CareerBuilder\footnote{\url{https://prn.to/3NugUWa}} 
reveals that 70\% of employers utilize social media as a screening tool for potential hires. Despite the potential utility or innocuous intentions perceived by recruiters, this practice raises concerns about potential discrimination against candidates and the disclosure of information forbidden during the interview process. For example, \citep{acquistiExperimentHiringDiscrimination2020} showed how the disclosure of certain personal information online can influence the hiring decisions of some U.S. employers, confirming the problematic nature of this practice.

\begin{figure}[ht]
    \begin{subfigure}{0.3\textwidth}
        \centering
        \includegraphics[width=\textwidth]{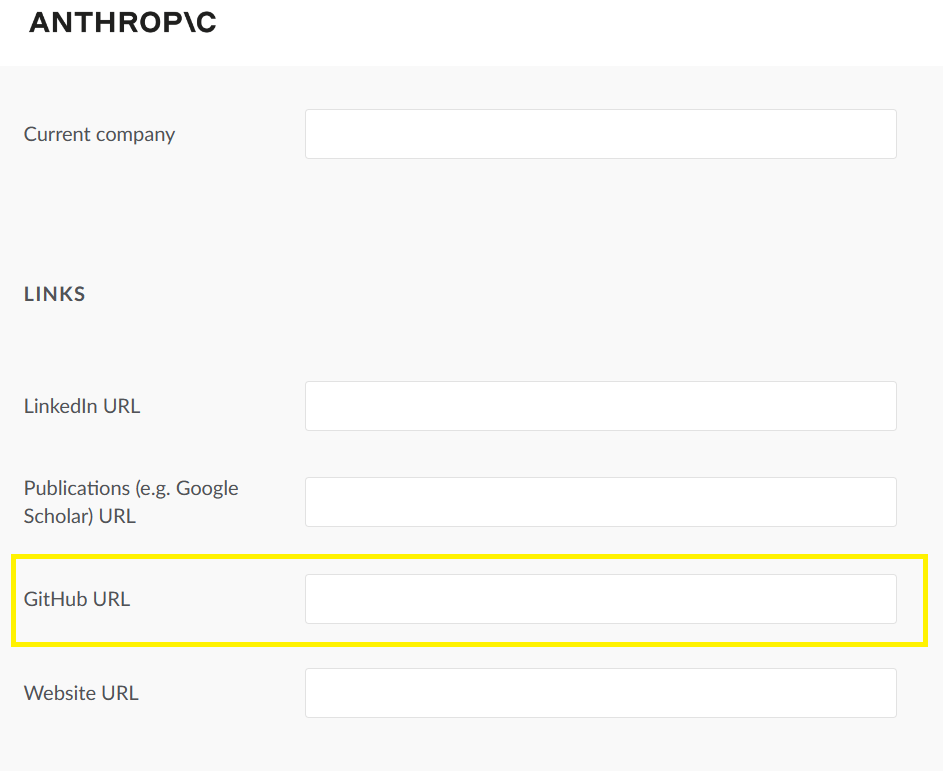}
        \caption{Online application for Anthropic.}
        \label{fig:anthropic}
    \end{subfigure}
    \hfill
    \begin{subfigure}{0.7\textwidth}
        \centering
        \includegraphics[width=\textwidth]{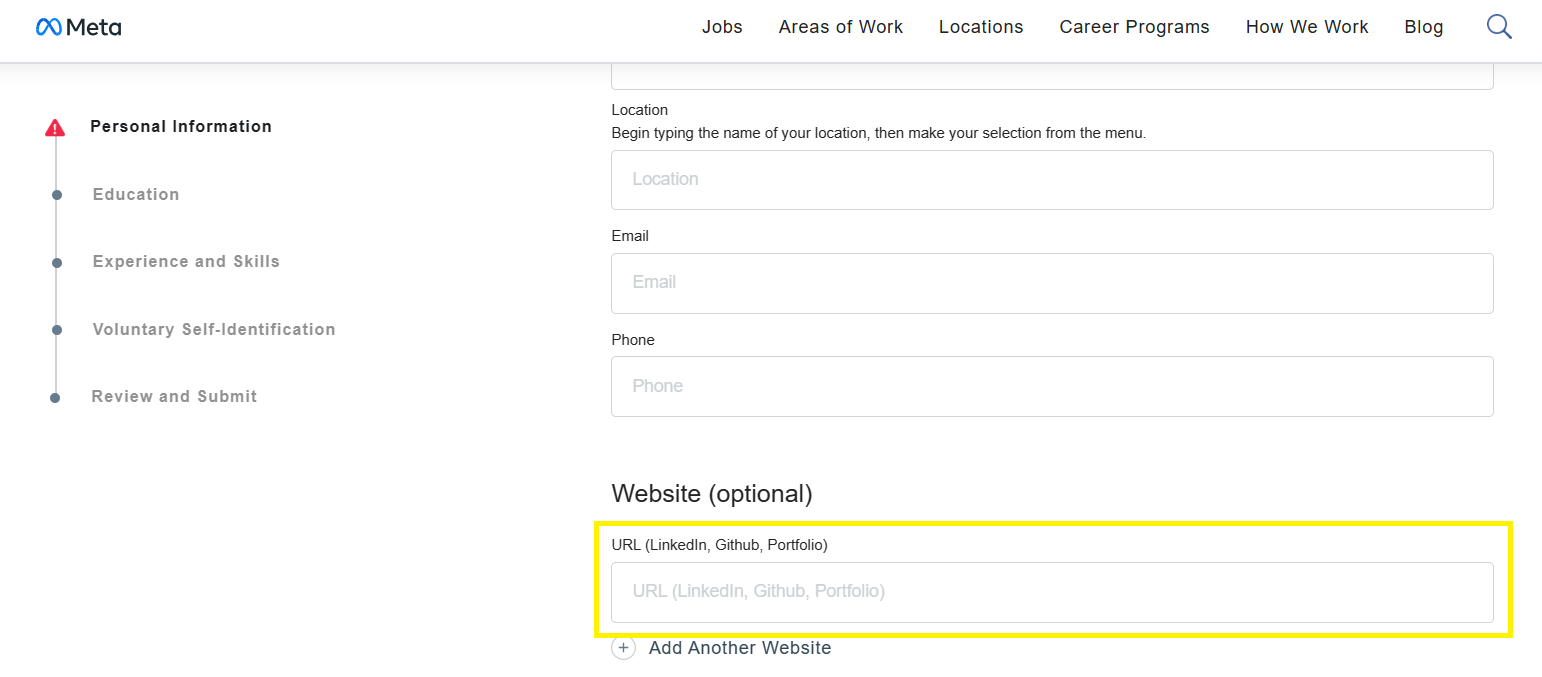}
        \caption{Online application for Meta.}
        \label{fig:meta}
    \end{subfigure}
\caption{Examples of companies requiring the \GH profile. }
\label{fig:application}
\end{figure}

Currently, it is a common practice for companies to request a candidate's \GH profile as part of the hiring process. Figure \ref{fig:application} displays a screenshot of the online application procedures for two renowned companies: Anthropic (Figure \ref{fig:anthropic}) and Meta (Figure \ref{fig:meta}). In addition to the conventional inclusion of LinkedIn profiles and links to Google Scholar publications, both companies demand the inclusion of the \GH URL.

The companies under consideration are technology giants, and requesting the \GH profile could constitute a component of the verification process for assessing technical skills. 
However, on \GH, numerous interactions among users have been scrutinized by various researchers to tackle and alleviate social issues such as toxicity or stress \citep{millerDidYouMiss2022,ramanStressBurnoutOpen2020}, showing that the usage of this platform goes beyond the sharing of technical knowledge.

Understanding the nature of information accessible on \GH is not only essential for the evidence we gathered regarding recruiter interest in this platform, but also from a broader perspective as it contributes to the overall “digital breadcrumbs" left by users \citep{strettonDangersOurTrail2015}. The significance lies not solely in the context of a single platform; rather, it pertains to the aggregation of information within the digital realm that can be associated with a user and exploited for different purposes, such as surveillance or psychological targeting \citep{lustgartenDigitalPrivacyMental2020,matzPrivacyAgePsychological2020,millerSurveillanceDigitalTrail2010}.

%% file: tex/RelatedWork.tex
This section reviews empirical studies on GitHub and findings on privacy issues on this platform. Moreover, we review works on the analysis of user-generated content and users' privacy. We focus on works that have produced labeled corpora as well as privacy profiling.

\subsection{\GH studies}
\GH is a widely used platform, and numerous researchers have conducted empirical studies on this platform. In their exploratory study,~\cite{henning2023understanding} examined reported issues related to data protection on \GH, shedding light on the influence of data protection regulations throughout the entire software development process.~\cite{acar2017security} conducted a study aimed at enhancing security decision-making for IT professionals. The researchers conducted an experiment with 307 active \GH users to assess their performance on security-related programming tasks. They found differences in performance related to experience levels, but no significant disparities based on student or professional status. ~\cite{10.1145/3510458.3513014} conducted an analysis of comments on \GH in order to uncover issues that are centred around human concerns. The analysis revealed a diverse array of issues, encompassing topics such as privacy and security.

\revised{Many researchers have focused on investigating a more social aspect of \GH.} 
~\cite{sajadi2023interpersonal} explored the dimension of interpersonal trust in Open Source Software (OSS) teams and how it is exhibited. The study analyzes 100 \GH pull requests from Apache Software Foundation projects to understand how trust is expressed in these interactions.~\cite{guzman2014sentiment} investigated the impact of emotions on productivity, task quality, creativity, group relationship, and job satisfaction through sentiment analysis of commit comments in various open-source projects. The results indicate that Java projects tend to receive more negative commit comments, while projects with widely distributed teams tend to have more positive emotional content. \cite{ramanStressBurnoutOpen2020} studied the problem of stress and burnout in open-source environments due to toxic discussions on GitHub issues. They demonstrated that a combination of pre-trained detectors for negative sentiment can effectively identify these issues. Furthermore, they established that classification accuracy is enhanced through domain adaptation. In their study, ~\cite{BLINCOE201630} defined popular users as individuals who provide guidance to OSS developers when they join new projects. The authors observed that those users did not possess the highest contribution rate. 
~\cite{10.1145/3510003.3510111} curated a sample of 100 toxic \GH issue discussions to gain an understanding of the characteristics of open-source toxicity. They found that some of the most prevalent forms of toxicity are entitled, demanding, and arrogant comments from project users as well as insults arising from technical disagreements. 

Different authors have investigated more specifically solely the topic of privacy leakages and self-disclosure on GitHub. According to~\cite{vasilescu2015data}, their user survey revealed that platform users are aware of certain demographic details about other developers, including gender, real names, and countries of residence. Additionally,~\cite{fordCodeItselfHow2019} demonstrated that GitHub developers explore a much wider array of information while scrutinizing pull requests, particularly information tied to the identity of the person submitting the pull request. This implies that GitHub not only contains identity information but also that such information is exploited during the evaluation of pull requests. In a similar vein,~\cite{meliHowBadCan2019} discovered that hundreds of thousands of API and cryptographic keys are leaked on GitHub at a rate of thousands per day. Meanwhile,~\cite{niuCodexLeaksPrivacyLeaks} demonstrated that it is possible to extract sensitive personal information from the Codex model used in GitHub Copilot.

From these studies, it is evident that GitHub has garnered attention from the software engineering community for several years. The primary findings indicate that various user information, including gender, real names, and countries, can be extracted from GitHub. Furthermore, the platform faces general privacy threats, such as the leakage of crypto-related secrets and private information on Copilot. Additionally, GitHub exhibits characteristics akin to other social networks, such as language toxicity, and is subject to study in the context of team working and cooperation. \revised{
It is worth noting 
that none of the mentioned studies investigated the adoption of privacy settings provided by the platform. Previous research has explored the adoption of privacy settings on other platforms such as Facebook \citep{fiesler2017or,chen2019demystifying,kanampiu2019privacy}, primarily through user surveys. To the best of our knowledge, this is the first study on GitHub privacy settings that is conducted based on the actual selections made by users.}

\subsection{Analysis of user-generated content}
 

User-generated content plays a significant and ubiquitous role on various platforms, attracting extensive attention from researchers for diverse purposes such as sentiment analysis, risk detection of depression (e.g., on Reddit), marketing analysis, and self-disclosure~\citep{SentimentAnalysisTourism,tadesseDetectionDepressionRelatedPosts2019,timoshenkoIdentifyingCustomerNeeds2019,umarDetectionAnalysisSelfDisclosure2019}. In this section, we present an overview of studies focused on the analysis of user-generated content, with a specific emphasis on the detection of self-disclosure.

\cite{bioglio2022analysis} conducted a study aimed at automatically detecting the sensitivity of Facebook posts. Their research involved analyzing a dataset comprising 9,917 Facebook posts, each one annotated by three experts as either sensitive, non-sensitive, or of unknown sensitivity. To enhance their investigation, two additional datasets were incorporated for comparative analysis. The first dataset consisted of posts extracted from Reddit, manually labeled to align with the Facebook corpus. The second dataset comprised anonymous posts from Whisper, considered sensitive, alongside non-sensitive tweets sourced from Twitter (now renamed as \textbf{X}).

In their experiment, the researchers employed four distinct classifiers on the datasets: Convolutional Neural Network (CNN), Recurrent Neural Network (RNN) with gated recurrent unit, RNN with long short-term memory, and BERT. Notably, they highlighted the limitations of existing corpora for sensitivity detection, emphasizing that many are often derived from specific topics, and thus incapable of detecting sensitivity on wider topics. Furthermore, their findings demonstrated that, on their datasets, RNNs and BERT exhibited significant performance in classification, and that lexical features are not sufficient for discriminating between sensitive and non-sensitive texts.

In their examination of social media safety, ~\cite{disalvoSocialMediaSafety2022a} introduced a methodology for categorizing social media posts by utilizing a pre-established corpus of violent phrases. Specifically focusing on Twitter, the researchers collected posts that were then organized using the labelTweets function. This function evaluates a tweet based on whether it contains a word from a predefined array of 'violent' terms, either accepting or rejecting the tweet accordingly. Subsequently, the authors manually assigned labels (negative, positive, or neutral) to the output generated by the labelTweets function, guided by corresponding corpora. The resulting corpus comprised nearly 600 tweets. The authors proceeded with a classification task to differentiate between negative and positive tweets, employing three classifiers: Naive Bayes Classifier, Support Vector Machine, and Logistic Regression. Their analysis concluded that the Naive Bayes Classifier performed the least effectively among the three.

~\cite{blosePrivacyCrisisStudy2020} conducted a study on self-disclosure patterns on social media, specifically focusing on tweets posted during the Coronavirus pandemic. The authors utilized a dataset comprising Tweet IDs, which was updated through the Amazon Web Service. Their methodology involved a pre-processing of the dataset, for filtering only content posted by individual users. To automate the detection of self-disclosure, the researchers employed a dictionary obtained from~\cite{umarDetectionAnalysisSelfDisclosure2019} and compared their dataset against an already annotated Twitter dataset. This comparison yielded satisfactory results, demonstrating the effectiveness of their approach. To better understand disclosure trends, the authors conducted a topic-modeling analysis on the described corpus. Furthermore, they conducted a comparative analysis, juxtaposing their observations of self-disclosure behaviors during the Coronavirus pandemic with those observed during Hurricane Harvey in 2017. The findings of the study revealed an increase in self-disclosure behaviors during the pandemic. In their conclusion, the authors encourage further research engagement to “better understand more subtle, voluntary self-privacy violations." This underscores the importance of ongoing investigations into evolving patterns of online self-disclosure, especially during critical events such as the Coronavirus pandemic.

Despite their distinct objectives, all these studies analyzing user-generated content share a common procedure, which can be summarized as follows: selecting a specific platform or online social network, retrieving textual data from this platform for preprocessing, and utilizing domain-specific resources like the Privacy Dictionary~\citep{gill2011privacy,vasalouPrivacyDictionaryNew2011b} or the self-disclosure dictionary~\citep{umarDetectionAnalysisSelfDisclosure2019} to support the preprocessing phase. Ultimately, classifiers are employed to automate the detection process.

In our research, we adhered to the same methodology described above for self-disclosure detection in pull\_requests comments. We focused on creating a multi-label corpus, where each comment is annotated for every sensitive information shared in the same text.

\subsection{User's privacy studies} 
With the increasing attention to the issue of privacy in the digital world, different researchers have investigated how to capture users' privacy desiderata and users' behaviours in online platforms.

\cite{brandao2022prediction} analyzed users' privacy profiles for what concerns mobile settings and exploited this information to predict the users’ answers to a permission request, while preserving user privacy by applying a federated learning approach.

\cite{tahaei2020understanding} examined Stack Overflow for privacy-related challenges faced by developers. They used topic modelling on 1,733 questions, identifying key topics. The results indicated that developers do seek support on privacy issues on Stack Overflow.

In an attempt to study privacy issues in Twitter (now \textbf{X})~\citep{kekulluoglu2020analysing}, the authors collected a dataset of 635k tweets containing the expression \emph{``happy for you.''}
By employing LDA topic modeling, the tweets were categorized into 12 distinct clusters representing different life events. They found out that around 8\% of the tweets mentioned protected users, with varying rates across different topics.

%% file: tex/methodology.tex

\begin{figure}[!htp]
    \centering
    \includegraphics[scale=0.5]{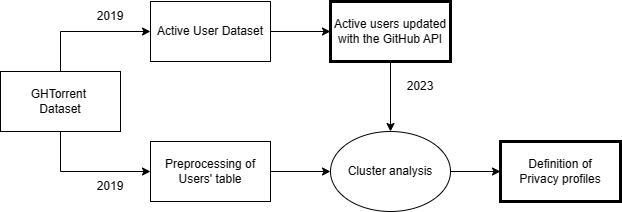}
    \caption{Workflow of the study.}
    \label{fig:workflow}
\end{figure}

A primary goal of this study is \revised{to investigate the use of privacy settings on \GH, that is, users' privacy desiderata expressed on the platform, and their potential limitations (RQ1)}. To this end, we started by exploring the largest existing dataset of GitHub activities, the GHTorrent dataset \citep{gousios2013ghtorent,gill2011privacy}. In this dataset, the \textit{Users} table provides the set of privacy settings selected by each \GH user \revised{as their privacy preferences on the platform. We adopted this information as being users' privacy desiderata.} Furthermore, from the GHTorrent dataset, we defined and selected \revised{a subset of users being more ``active'' on \GH, in order to conduct a more fine-grained analysis and to have more textual data to analyze. We called this dataset the \textit{Active users}. We 
updated the \textit{Active users} dataset due to the absence of certain privacy settings that were missing from the original GHTorrent dataset, such as email addresses and social media channels. This process is detailed in Section \ref{subsec:data_curation}.} By strategically retrieving additional data from the current database, it was possible to effectively address the gaps in users' metadata and ensure a thorough understanding of user data. This was achieved through the use of GitHub APIs\footnote{\url{https://docs.github.com/en/rest?apiVersion=2022-11-28}} \revised{and the users' login information,} which allowed for the seamless integration and enrichment of GHTorrent existing data with detailed and updated user-specific insights. 
On both datasets (\textit{Users} and \textit{Active users}), we conducted a cluster analysis to define users' privacy profiles and \revised{observe differences between users selection of settings}. All these steps are detailed in Sections \ref{subsec:data_curation}, 
\ref{subsec:cluster_analysis}, and represented in Figure \ref{fig:workflow}.

To investigate RQ2 and RQ3, \revised{that is, the information disclosed on the platform and the analysis of users' behavior}, we considered textual comments provided by the GHTorrent dataset. In particular, we analyzed the pull\_request comments to find privacy-sensitive information disclosed intentionally or unintentionally by the users. The methodology for this analysis is described in Section \ref{subsec:corpus}. In addition, it should be noted that GHTorrent truncates textual comments, specifically those found in commits, issues, and pull\_requests. In order to facilitate the empirical investigation and analysis of privacy awareness in pull\_requests, we acquired the original text through the use of \GH APIs.

\begin{table*}[t!]
\centering
\scriptsize
\caption{Table \textit{Users} from the GHTorrent dataset.}
\begin{adjustbox}{max width=\textwidth}
\begin{tabular}{p{1.2cm}p{1.2cm}p{1.2cm}lllp{1cm}p{1.2cm}p{0.5cm}p{1cm}p{1.7cm}lp{0.3}}

\toprule
        \textbf{login} &       \textbf{company} &           \textbf{created\_at} & \textbf{type} &  \textbf{fake} &  \textbf{deleted} &          \textbf{long} & \textbf{lat} & \textbf{country code} &                                           \textbf{state} &          \textbf{city} &  \textbf{location} \\
\midrule
     $U_{1}$ \hl{hid} &            -- &  2016-04-18 11:42:46 &  USR &     0 &        1 &            -- &            -- &           -- &                                              -- &            -- &                          -- \\
       $U_{2}$ \hl{hid} &     Sage GmbH &  2008-12-15 12:28:33 &  USR &     0 &        0 &    0.00000000 &    0.00000000 &           -- &                                              -- &            -- &            Rastede, Germany \\
     $U_{3}$ \hl{hid} &            -- &  2008-03-22 00:37:42 &  USR &     0 &        0 &  132.45529270 &   34.38520290 &           jp &                            Hiroshima Prefecture &     Hiroshima &                   Hiroshima \\
     $U_{4}$ \hl{hid} &            -- &  2012-08-03 16:08:15 &  USR &     1 &        1 &            -- &            -- &           -- &                                              -- &            -- &                          -- \\
  $U_{5}$ \hl{hid} &  @ValuationUp &  2008-04-28 17:25:53 &  USR &     0 &        0 &   28.04730510 &  -26.20410280 &           za &  City of Johannesburg &  Johannesburg &  Johannesburg, South Africa \\

 $\cdots$ & $\cdots$& $\cdots$& $\cdots$& $\cdots$& $\cdots$& $\cdots$& $\cdots$& $\cdots$& $\cdots$& $\cdots$ & $\cdots$\\
\bottomrule
\end{tabular}
\end{adjustbox}

\label{tab:users}
\end{table*}

\subsection{Data curation}\label{subsec:data_curation}
When talking about users' private information, we refer to the personal information they do not want to share. On GitHub, these \textit{privacy desiderata} can be expressed through the profile settings where the user chooses to reveal or hide certain information (company, location, bio, social accounts, and so on). In the GHTorrent dataset \citep{gousios2013ghtorent,gill2011privacy}, this information was collected into the \textit{Users}' table (see Table \ref{tab:users}).  \revised{For each user in the GHTorrent dataset, we were able to retrieve information regarding whether the user chose to share the name of their company on their profile, along with details about their location, including city and state. These choices reflect the decisions made by users at the time of the last GHTorrent updates, which occurred in 2019. To overcome these limitations, we updated the privacy choices of the \textit{Active users} dataset.} The data curation process is summarized in Table \ref{tab:datacuration}. 
\begin{table}[h!]
\scriptsize
\centering
\caption{Data curation process.}
\begin{tabular}{cp{1.8cm}p{1.2cm}p{2.3cm}}
\toprule
\textbf{Step} & \textbf{Source} & \textbf{\#Users}  &\textbf{\#Pr Comments} \\
\midrule
\par\vspace{5pt}
Step 1 & GHTorrent  & 22,525,012 & 38,159,840 \\
\par\vspace{5pt}
Step 2 & \textit{Active users} & \revised{6,329} & 89,749 \\
\par\vspace{5pt}
Step 3 & \GH API actualization & 6,132  &  15,672   \\
\bottomrule
\end{tabular}
\label{tab:datacuration}
\end{table}

\paragraph{Step 1: GHTorrent preprocessing Users}
We used the GHTorrent dataset to conduct a cluster analysis on a broader population of \GH users. 

\begin{figure}[h!]
        \vspace{-.3cm}
        \centering
        \includegraphics[width=.80\linewidth]{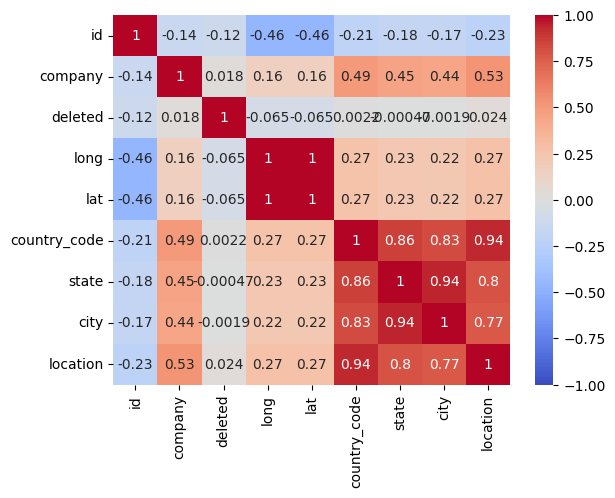}
        \vspace{-.3cm}
        \caption{Correlation matrix \revised{of the variables in the \textit{Users} dataset}.}
        \label{fig:correlation}
        \vspace{-.2cm}
\end{figure}

The \textit{Users'} table from the GHTorrent dataset contains 32,411,628 accounts, from which we eliminated those marked as \texttt{organization}, \texttt{fake} and \texttt{deleted}. This resulted in a set of 22,525,012 users. \revised{Figure \ref{fig:correlation} shows the correlation matrix of the variables in \textit{Users} table, obtained using the Python library Scikit-learn. Based on this correlation matrix, we excluded the \texttt{state}, \texttt{country code}, \texttt{lat}, and \texttt{location} variables because of the value of correlation being very high (above 0.7), following the methodology outlined by  \cite{king2015cluster}}. 

Every column of the dataset was converted into 0 if the value was missing, and 1 if the value was present. This choice was made because we were only interested in evaluating whether a user would disclose (1) or not (0) that piece of information. The resulting dataset is represented in Table \ref{tab:userpre}. In 
the tables we hide information related to user identification using the \hl{Hid} field, due to privacy reasons.

\begin{table}[h]
\footnotesize
\centering
\caption{\textit{Users}' table preprocessed.}
\begin{tabular}{rccc}
\toprule
 \textbf{login} & \textbf{company} &  \textbf{long} & \textbf{city} \\
\midrule
 $U_{1}$ \hl{hid} & 1 &  1 & 0 \\
 $U_{2}$ \hl{hid} & 0 &  1 & 1 \\
 $U_{3}$ \hl{hid} & 0 &  0 & 0 \\
 $U_{4}$ \hl{hid} & 1 &  1 & 1 \\
 $U_{5}$ \hl{hid} & 1 &  1 & 1 \\
 $\cdots$ &  $\cdots$ & $\cdots$ & $\cdots$ \\
\bottomrule
\end{tabular}
\label{tab:userpre}
\end{table}

\paragraph{Step 2: Active users}

To have a more updated version of  \revised{users' privacy preferences}, we restricted our research to those users considered more ``active'' on \GH, those who may be identified by quantifying the number of actions they have performed on the platform. Therefore, we constructed a new dataset that contained information about the number of \CO, \CC, \FO, \PRC and \IC executed by each user (see Table \ref{tab:aggregate}). This achievement was made possible through the GHTorrent dataset since this dataset allows the extraction of the precise count for each action executed by the users. Indeed, we retrieved this information by the corresponding tables from the GHTorrent dataset. After the due preprocessing step, we performed a cluster analysis on the dataset described in Table \ref{tab:aggregate}. Primarily, we scaled this dataset, and performed a K-means cluster analysis on the dataset with a number of clusters K=4 (found with the Elbow method). 

\begin{table}[h]
\centering
\scriptsize
\caption{Dataset of actions performed by each user on \GH. \revised{On this dataset, we performed a cluster analysis to define the more ``active" users on the platform. }}
\begin{tabular}{lp{1.5cm}p{1.5cm}p{1.5cm}p{1.5cm}p{1.5cm}}
\toprule
\textbf{user\_id} & \textbf{\#pull request comments} & \textbf{\#followers} & \textbf{\#commits} & \textbf{\#commit comments} & \textbf{ \#issue comments} \\
\midrule
0 & 11.0 & -- & -- & 14.0 & -- \\
1 & 15.0 & 31.0 & 21.0 & 6.0 & 30.0 \\
2 & 25.0 & 400.0 & 3598.0 & 25.0 & 1504.0 \\
3 & 4.0 & -- & 2.0 & 3.0 & -- \\
4 & 5.0 & 174.0 & 463.0 & 14.0 & 516.0 \\

$\cdots$ & $\cdots$ & $\cdots$& $\cdots$& $\cdots$& $\cdots$ \\
\bottomrule
\end{tabular}
\label{tab:aggregate}

\end{table}

Secondly, we discretized the variables into three distinct bins. The ``low" label was assigned to instances where the number of actions ranged from the minimum value to the 65th percentile. The ``medium" label was applied when the number of actions fell between the 65th percentile and the mean value. Finally, the ``high" label was assigned to cases where the number of actions varied from the mean value to the maximum value. The choice of the 65th percentile as a threshold was deliberate, as it consistently yielded values lower than the mean across all variables in the dataset, as evidenced by the statistics in Table \ref{tab:statistics}. 

\begin{table}[h]
\scriptsize
\centering
 \caption{Statistics from the dataset of actions performed by users on GitHub.}
\begin{tabular}{lp{2cm}p{1.5cm}p{1.5cm}p{1.5cm}p{1.5cm}}
\toprule
{} &  \textbf{\#pull request comments} & \textbf{\#followers} & \textbf{\#commits} & \textbf{\#commit comments} & \textbf{\#issue comments}\\
\midrule
\textbf{Count} & $1.2 \times 10^5$ & $1.2 \times 10^5$ & $1.2 \times 10^5$ & $1.2 \times 10^5$ & $1.2 \times 10^5$ \\
      \textbf{Mean}  & $4.9 \times 10^{-4}$ & $1.5 \times 10^{-3}$ & $3.9 \times 10^{-3}$ & $2.5 \times 10^{-3}$ & $4.9 \times 10^{-3}$ \\
        \textbf{Std}   & $4.4 \times 10^{-3}$ & $9.7 \times 10^{-3}$ & $1.2 \times 10^{-2}$ & $1.2 \times 10^{-2}$ & $1.5 \times 10^{-2}$ \\
       \textbf{Min}   & $0.0 \times 10^0$ & $0.0 \times 10^0$ & $0.0 \times 10^0$ & $0.0 \times 10^0$ & $0.0 \times 10^0$ \\
        \textbf{25\%}   & $0.0 \times 10^0$ & $1.3 \times 10^{-4}$ & $7.9 \times 10^{-5}$ & $2.5 \times 10^{-4}$ & $3.0 \times 10^{-4}$ \\
        \textbf{50\%}   & $0.0 \times 10^0$ & $3.7 \times 10^{-4}$ & $7.1 \times 10^{-4}$ & $5.1 \times 10^{-4}$ & $1.0 \times 10^{-3}$ \\
        \textbf{65\%}   & $3.3 \times 10^{-5}$ & $6.1 \times 10^{-4}$ & $1.8 \times 10^{-3}$ & $1.0 \times 10^{-3}$ & $2.1 \times 10^{-3}$ \\
        \textbf{75\%}   & $9.8 \times 10^{-5}$ & $9.3 \times 10^{-4}$ & $3.3 \times 10^{-3}$ & $1.7 \times 10^{-3}$ & $3.5 \times 10^{-3}$ \\
       \textbf{Max}   & $1.0 \times 10^0$ & $1.0 \times 10^0$ & $1.0 \times 10^0$ & $1.0 \times 10^0$ & $1.0 \times 10^0$ \\
\bottomrule
\end{tabular}
\label{tab:statistics}
\end{table}

\begin{figure}[h!]
    \begin{subfigure}{0.40\linewidth}
        \centering
        \includegraphics[width=\linewidth]{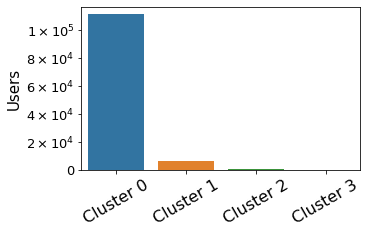}
        \caption{Users per cluster}
        \label{fig:active}
    \end{subfigure}
    \hfill
      \begin{subfigure}{0.45\linewidth}
        \centering
        \includegraphics[width=\linewidth]{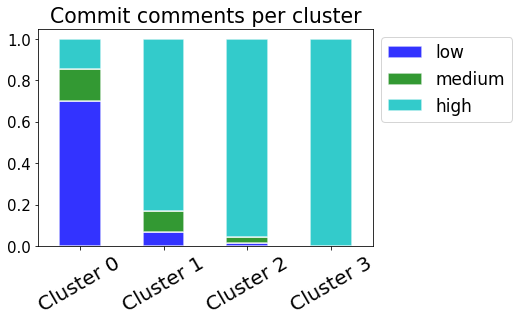}
        \caption{Commit comments}
        \label{fig:commit}
    \end{subfigure}
    \hfill
     \par\bigskip
    \begin{subfigure}{0.45\linewidth}
        \centering
        \includegraphics[width=\linewidth]{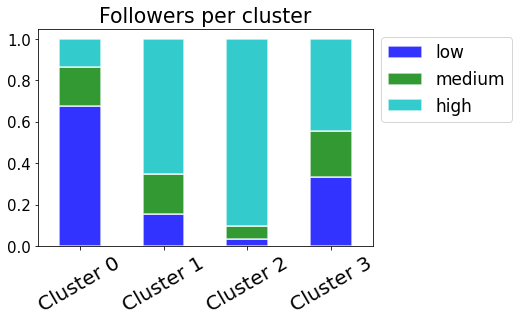}
        \caption{Followers}
        \label{fig:followers}
    \end{subfigure}
    \hfill
    \begin{subfigure}{0.45\linewidth}
        \centering
        \includegraphics[width=\linewidth]{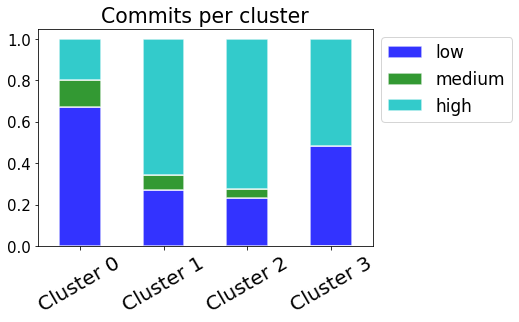}
        \caption{Commits}
        \label{fig:commits}
    \end{subfigure}
    \hfill
    \par\bigskip
    \begin{subfigure}{0.45\linewidth}
        \centering
        \includegraphics[width=\linewidth]{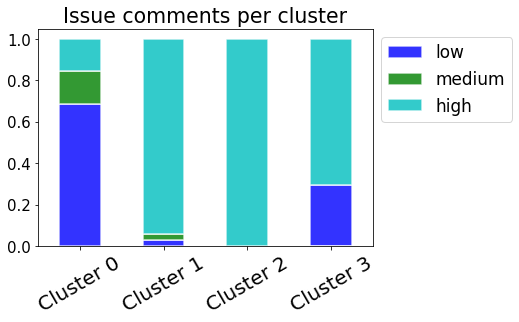}
        \caption{Issue comments}
        \label{fig:issues}
    \end{subfigure}
    \hfill
    \begin{subfigure}{0.45\linewidth}
        \centering
        \includegraphics[width=\linewidth]{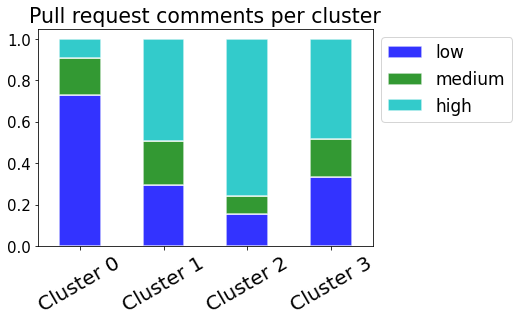}
        \caption{Pull requests comments}
        \label{fig:pullrequest}
    \end{subfigure}
    
    \caption{\revised{Cluster analysis on the dataset of the actions performed by each user.} Figure \ref{fig:active} shows the number of users per cluster, Figures \ref{fig:commit}, \ref{fig:followers}, \ref{fig:commits}, and \ref{fig:issues} show the distribution of variables per cluster.}
   \label{fig:cluster_active}
\end{figure}

Figure \ref{fig:active} shows the different clusters of users, according to the number of actions performed. Figures \ref{fig:commit}, \ref{fig:followers}, \ref{fig:commits}, and \ref{fig:issues} illustrate the percentage of each variable per cluster. For instance, examining Figure \ref{fig:commit} reveals that users in Cluster 0 tend to make few commit comments (the variable 'low' is predominant in this cluster).
The analysis of the variables per cluster in Figure \ref{fig:cluster_active} allowed us to clearly identify the \textit{least} active users, being cluster 0. \revised{Consequently, we excluded this cluster of  \textit{Non-Active users} from the overall dataset, resulting in what we refer to as the \textit{Active users} dataset (6,329 users in total).} \revised{The number of \textit{Active users} decreased to 6,132 with the updates of users' privacy settings that we have performed through the \GH API, as explained in the next step. This is attributed to users' departing the platform by 2023. }

\paragraph{Step 3: Updating active users privacy settings} Once we selected the \textit{Active users}, we updated their privacy settings choices using the GitHub API \revised{with their login}. \revised{This action was taken due to the incomplete nature of the GHTorrent dataset, which lacked information such as users displaying their email addresses, events, and their Twitter accounts.} The updated dataset is illustrated in Table \ref{tab:active users}. Compared with the GHTorrent dataset, the columns added were \texttt{email}, \texttt{events} and \texttt{twitter}. Some of the users in the \textit{Active users} dataset were no longer present on \GH \revised{at the time of the update} so they were eliminated. The final number of updated \textit{Active users} is \numDevs users. The methodology detailed in this section is represented in Figure~\ref{fig:workflow}.

\begin{table}
\centering

\caption{The \textit{Active users} dataset updated through the GitHub API.}
\rotatebox{90}{
\begin{tabular}{p{1cm}p{1.4cm}p{1.5cm}p{1.5cm}p{1.3cm}p{1.3cm}p{1cm}p{1.5cm}p{1cm}p{1cm}p{1cm}lp{0.5cm}}

\toprule
 \textbf{login} & \textbf{company} & \textbf{created\_at} & \textbf{location} & \textbf{long} & \textbf{lat} & \textbf{country code} & \textbf{state }& \textbf{city} & \textbf{email} & \textbf{events} & \textbf{Twitter} \\
\midrule
 $U_{1}$ \hl{hid} & @floraison & 2008-03-22 00:37:42 & Hiroshima & 132.45529270 & 34.38520290 & jp & Hiroshima Prefecture & Hiroshima & $E_{1}$ \hl{hid} & 151 & -- \\
 $U_{2}$ \hl{hid} & DNSimple & 2008-04-06 08:44:35 & Rome, Italy & 12.49636550 & 41.90278350 & it & Rome & Rome & $E_{2}$ \hl{hid} & 161 & $T_{2}$ \hl{hid} \\
 $U_{3}$ \hl{hid} & -- & 2010-02-05 06:35:04 & -- & 0.00000000 & 0.00000000 & -- & -- & -- & -- & 30 & -- \\
 $U_{4}$ \hl{hid} & @platformatic & 2009-02-05 21:24:19 & Forlì, Italy & 0.00000000 & 0.00000000 & -- & -- & -- & $E_{4}$ \hl{hid} & 291 & $T_{4}$ \hl{hid} \\
 $U_{5}$ \hl{hid} & -- & 2011-01-23 18:48:07 & -- & 0.00000000 & 0.00000000 & -- & -- & -- & $E_{5}$ \hl{hid} & 284 & -- \\
 $U_{6}$ \hl{hid} & Cocos & 2010-11-19 08:06:45 & Xiamen Fujian China & 118.08942500 & 24.47983300 & cn & Fujian & Xiamen & $E_{6}$ \hl{hid} & 290 & $T_{6}$ \hl{hid} \\
\bottomrule
\end{tabular}
}
\label{tab:active users}
\end{table}
After the data curation described in Section \ref{subsec:data_curation}, we conducted a cluster analysis on the two datasets: \textit{Users} and \textit{Active users}. The goal was to understand whether users present variability in terms of privacy settings on GitHub, \ie if there are different \textit{privacy profiles} that can be associated with 
the users of this platform \citep{diruscio2022} \revised{meaning that users' adopt different privacy settings as a tool for their privacy}. The clustering techniques adopted were K-means and hierarchical clustering, as suggested by different authors \citep{sanchez2020recommendation,brandao2022prediction}. 

\subsection{Cluster analysis}\label{subsec:cluster_analysis}
To define the privacy profiles of \GH users, we performed cluster analysis on the two datasets, \textit{Users} and \textit{Active users}. The variables considered are the privacy settings chosen by each user, however, there are variations between the two sets due to recent additions of privacy features by \GH, such as \revised{\texttt{events} and \texttt{twitter}.} This novelty is reflected in the updated dataset of \textit{Active users}. \revised{In the \textit{Active users} dataset, we have added the \texttt{email} field, which was removed from the \textit{Users} dataset by the author of the GHTorrent dataset.}

For the \textit{Users} dataset preprocessed as described in Section \ref{subsec:data_curation} \textit{Step 1}, we performed a K-means cluster analysis~\citep{hartigan1979algorithm} with Euclidean distance. The number of clusters K=3 was chosen with the Elbow method \citep{syakur2018integration}. 

\begin{figure}
        \centering
        \includegraphics[width=.8\linewidth]{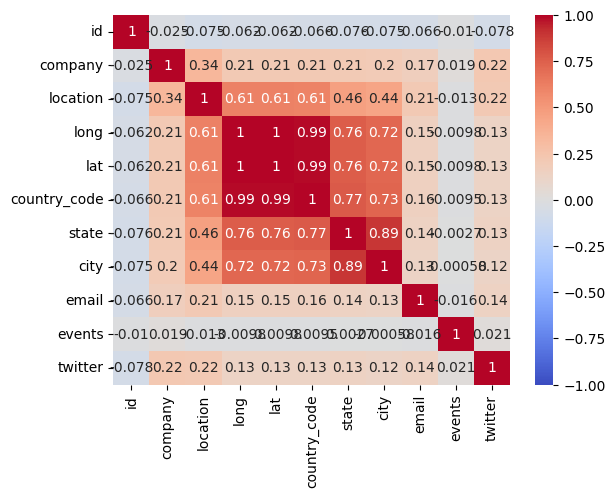}
        \caption{Correlation matrix of the variables present in the \textit{Active users} dataset.}
        \label{fig:corr_active}
\end{figure}

\begin{figure}
\centering
    \includegraphics[width=.68\linewidth]{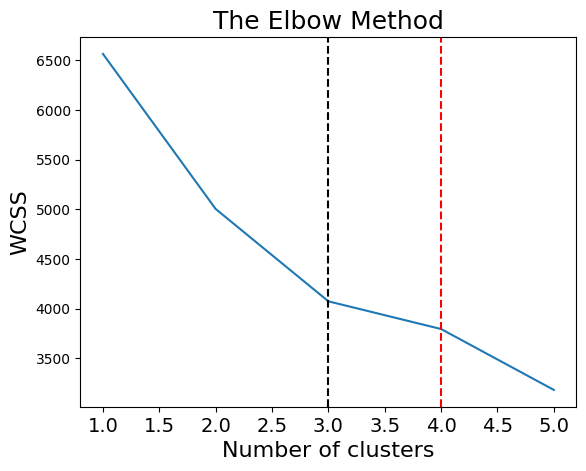}
    \caption{Elbow method on the \textit{Active users} dataset to establish the more appropriate number of clusters.}
    \label{fig:elbow_active}
    
\end{figure}

\begin{figure}
\centering
        \includegraphics[width=.6\linewidth]{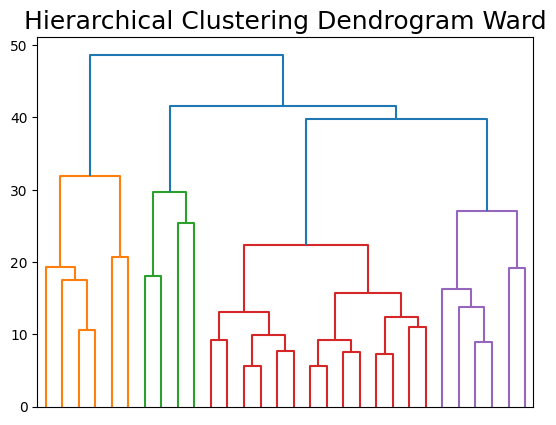}
        \caption{Dendrogram with Ward's method on the \textit{Active users} dataset to have an overview of the dataset and to confirm the appropriate number of clusters. }
        \label{fig:ward_dendro}
 \end{figure}

For the \textit{Active users} dataset updated as described in Section \ref{subsec:data_curation} \textit{Step 3}, we removed the variables \texttt{location}, \texttt{long}, \texttt{lat}, \texttt{country code}, and \texttt{state} from the \textit{Active users} dataset due to their correlations as shown by the correlation matrix in Figure \ref{fig:corr_active}. 
We employed both the Elbow method (Figure \ref{fig:elbow_active}) and an analysis of the dendrogram obtained using Ward's method (Figure \ref{fig:ward_dendro}) to determine an appropriate number of clusters. We identified K=4 as a reasonable number of clusters for this dataset, therefore a K-means cluster analysis was performed with this value.

Both cluster analysis for the datasets \textit{Users} and \textit{Active users} were conducted in order to study users' privacy profiles and to verify whether these settings were actively adopted by the users. The results are discussed in Section \ref{sec:EmpiricalResults}.


\subsection{Construction of the corpus}\label{subsec:corpus}
To address RQ2 and RQ3, we started exploring users' privacy behaviour for what concerns textual data, e.g., their comments on GitHub.  While there's a wealth of textual data in the GHTorrent dataset, we focused our research on pull\_requests comments exclusively (86,000 comments overall).
\revised{This decision was guided by previous studies highlighting a greater probability of encountering significant user interactions and consequently more sensitive information in pull\_request contexts \citep{sajadi2023interpersonal, iyer2019effects}.}
To collect the more privacy-sensitive data \revised{and prepare the dataset for subsequent manual labeling}, we automatically labeled each comment using the Privacy Dictionary created by \cite{vasalouPrivacyDictionaryNew2011b,gill2011privacy}, exploiting 
libraries provided by existing work \citep{casillo2022detecting}. This dictionary was constructed and validated by its authors through interviews and focus groups from different privacy-sensitive (offline and online) contexts, \revised{leading to identify eight different privacy categories, as illustrated in Table \ref{tab:categories}. Each category represents a distinct privacy realm, potentially encompassing different types of private information. Previous authors have successfully adopted this dictionary to detect privacy language patterns within a given text, such as \cite{bioglio2022analysis} and \cite{d2021most}. The final goal of this process is to gather evidence of self-disclosure on GitHub. 
} An example from the corpus labeled with the Privacy Dictionary is shown in Table \ref{tab:corpus}. \revised{The first column indicates the user who made the comment in the ``body" column. The ``Categories" column shows the privacy category assigned to each comment, identified through the ``Keywords" in the corresponding column. }

\begin{table}[ht]

\footnotesize
\centering
\caption{\revised{The eight privacy categories as described in the Privacy Dictionary.}}
\begin{tabular}{p{2cm}p{8.8cm}}
\toprule
\textbf{Category} & \textbf{Definition} \\
\midrule
NegativePrivacy & This category captures the antecedents and consequences of breaches in privacy. It encompasses terms associated with privacy concerns and risks as well as judgments about the source and type of violation. \\ \hline

NormsRequisites & NormsRequisites includes the norms, beliefs, and expectations in relation to achieving privacy. \\ \hline

OutcomeState & It includes words that describe the static behavioral states and the outcomes that are served through privacy. This grouping corresponds to Westin's (1967) delineation of privacy states and purposes. \\ \hline

PrivateSecret & It includes descriptors or terms that articulate the essence of privacy. This category aids in discerning precisely which elements individuals perceive as private. \\ \hline

Intimacy & Intimacy comprises words that portray and measure different facets of small-group privacy. It includes words that denote the psychological needs involved in revealing oneself to another person, as well as the emotional proximity that forms between individuals. \\ \hline

Law & This category includes words employed to describe legal definitions of privacy. \\ \hline
Restriction & Words in this category express the closed, restrictive, and regulatory behaviors employed in maintaining privacy. Thus, the Restriction category can be used to measure the behaviors that people take to protect their privacy. \\ \hline

OpenVisible & This category includes words that represent the dialectic openness of privacy. \\ \hline
\end{tabular}
\label{tab:categories}
\end{table}

\begin{table}[ht]
\footnotesize
\centering
\caption{Corpus labeled using the Privacy Dictionary.}
\begin{tabular}{p{0.9cm}p{2.4cm}p{0.8cm}p{1.8cm}p{1.5cm}}
\toprule
\textbf{user\_id} &                                               \textbf{body} &      \textbf{login} &                \textbf{Categories} &              \textbf{Keywords} \\
\midrule
     2 &  Hello Jun, I would prefer if the conversation ... &  $U_{1}$ \hl{hid} &      Intimacy OpenVisible &      ['conversation'] \\
      2 &  Could you please open an issue for that?Thanks... &  $U_{1}$ \hl{hid} &               OpenVisible &              ['open'] \\
       5 &  Makes sense. I'll hold off this change to a se... &     $U_{2}$ \hl{hid} &              OutcomeState &          ['separate'] \\
       5 &  FYI, I started a prototype long time ago:https... &     $U_{2}$ \hl{hid} &             PrivateSecret &          ['identity'] \\
      10 &  ahaha! I think it was a test that it did not w... &   $U_{4}$ \hl{hid} &  OutcomeState Restriction &  ['safely', 'delete'] \\
      10 &  this whole block on code should be indented of... &   $U_{4}$ \hl{hid} &               Restriction &             ['block'] \\
      10 &  sure! I'm kind of full at the moment, but I'll... &   $U_{4}$ \hl{hid} &               OpenVisible &            ['report'] \\
      10 &  This is not completely correct, as calling `th... &   $U_{4}$ \hl{hid} &              OutcomeState &           ['prevent'] \\
      10 &  Can you please release this? It's completely o... &   $U_{4}$ \hl{hid} &  OpenVisible OutcomeState &           ['release'] \\
      10 &  no idea. We have that code running in aws-ami-... &   $U_{4}$ \hl{hid} &              OutcomeState &              ['safe'] \\

$\cdots$ & $\cdots$ & $\cdots$ & $\cdots$ & $\cdots$ \\
\bottomrule
\end{tabular}
	\label{tab:corpus}
\end{table}

\begin{figure}[ht]
    \centering
    \includegraphics[width=\linewidth]{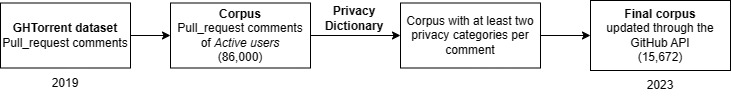}
    \caption{Process for constructing the corpus.}
    \label{fig:CorpusProcess}
\end{figure}

Given that GitHub pull\_request comments often involve technical details, we aimed to enhance the efficiency of identifying comments with private information. To achieve this, we specifically focus on comments that carry more than one label from the categories within the Privacy Dictionary. \revised{Indeed, due to the broad scope of the dictionary, we have empirically noted that most comments were assigned at least one label. Therefore, we chose to select comments with multiple labels that potentially included information from various privacy domains represented by each privacy category of the dictionary. This approach increases the likelihood of discovering more sensitive information, potentially different from what was discovered by previous authors \citep{vasilescuGenderTenureDiversity2015}.}
This selection resulted in a final corpus of 15,672 comments. However, since these comments in the GHTorrent dataset were truncated, we updated all of them through the GitHub API. This process in represented in Figure \ref{fig:CorpusProcess}.
\revised{This corpus, developed with the aid of the Privacy Dictionary, was manually labeled as described in the following Section \ref{subsec:manual_labeling} to classify sensitive comments and the type of information disclosed.}

\subsection{Manual labeling process and protocol}\label{subsec:manual_labeling}

After skimming the pull\_requests comments corpus, as outlined in Section \ref{subsec:corpus}, we curated a set of 2,000 comments, representing nearly 10\% of the extracted comments.  To ensure a sufficient level of informativeness, we specifically chose comments with a minimum of 2,000 characters. The aim was to explore the nature of information that users potentially disclose in their pull\_request comments. The selection of longer comments, coupled with their prior labeling using the Privacy Dictionary, aimed to enhance the likelihood of filtering out irrelevant or purely technical comments. These comments were subjected to manual labeling by all the authors of this paper.

The annotation team consisted of four members, ensuring a gender balance. All annotators had STEM backgrounds and held various academic positions, ranging from PhD student to full professor. Three of the annotators were from the same country, while the fourth was from a different one. Each annotator was given a file of roughly 1,000 comments; every file was assigned to
at least two different annotators, following existing guidelines about how to conduct a user study~\citep{DBLP:journals/ese/RoccoRSNR21,DBLP:journals/software/RobillardWZ10}. 
A value of 1 was assigned to comments that revealed personal information about the user, 0 otherwise, irrespective of our 
perception of sensitivity. Sensitivity is a subjective concept influenced by social and cultural factors; thus, what one deems sensitive may differ from person to person. Indeed, \revised{for the manual labeling process}, we refer to the definition provided by \cite{bioglio2022analysis} about Privacy-sensitive content: 

\begin{quote}
    
A generic user-generated content is privacy-sensitive if it discloses, explicitly or implicitly, any kind of personal information about its author or other identifiable persons. 
\end{quote}

For comments deemed as disclosing information, annotators had to select a label from the \textbf{Possible category} column corresponding to the type of information disclosed. The proposed labels were \textbf{Personal name}, \textbf{Workplace}, \textbf{Email}, \textbf{Location}, and \textbf{Gender}. The labels regarding personal name, job's information, email and location were derived from a preliminary corpus analysis conducted by one of the authors. Furthermore, these labels aligned with the privacy settings available on the GitHub profile, where users have the option to conceal specific private information. The inclusion of the gender label was prompted by findings from various studies on GitHub that highlight instances of gender discrimination or non-inclusive behaviors
\citep{imtiazInvestigatingEffectsGender2019,garciaUnderstandingOpenSource2023}. 
Annotators were allowed to choose multiple labels for each comment and could use the \textbf{Other} column to indicate information not covered by provided labels. Participants were given an annotation guide to enhance consistency throughout the process.

After completion, the four annotators met together via Teams in two different days to discuss the comments marked as privacy sensitive and to reach an agreement in case a comment was marked only once. Each comment expressing disagreement underwent thorough discussion between the two assigned annotators, while the remaining two played a moderating role, facilitating the conversation and aiding in reaching a consensus. \revised{During these discussions, one annotator was prompted to explain why a comment was considered sensitive or not sensitive to the other annotator of the same comment. Agreement would result in either flagging the comment as sensitive or deleting it. Moreover, we discovered during these meetings that several sensitive comments disclosed information not covered by the existing labels. This information was added by the annotators in the column \textbf{Other}. We unanimously decided to introduce additional labels, such as \textbf{Moral values}, \textbf{Community etiquette}, \textbf{Personal info}, \textbf{Language}, and \textbf{Relation with a user}, to account for instances where comments disclosed such information.} 
Ultimately, 147 comments were unanimously identified as disclosing private user information. Table \ref{tab: CommentsLabels}
provides an excerpt from the fully annotated corpus, displaying comments with specific labels and brief descriptions of the label meanings.

On the labeled corpus, we conducted an analysis to calculate the level of agreement. Each comment was annotated by two annotators, resulting in two raters per comment. Across the entire dataset, we computed Cohen's kappa coefficient, a metric utilized to assess inter-rater reliability \citep{cohen1960coefficient}. \revised{
The Kappa score of 0.49 was computed for the binary sensitivity column before any discussion occurred, indicating a moderate level of agreement among the raters} \citep{warrens2015five}. This is not surprising considering the novelty of the analyzed data and the nuances present in the dataset.  It could also be attributed to the observation that, out of 2,000 comments, 1846 received identical binary labels from both evaluators.

\FloatBarrier
\begin{table}[ht]

\caption{Examples of labels assigned to a comment and description of the label's meaning.}
\adjustbox{max width=\textwidth}
{

\begin{tabular}{p{9 cm}p{1.2 cm}p{3 cm}}
\hline
\textbf{Comments}  & \textbf{Label}      & \textbf{Description of label}   \\
\hline
1) I tried looking for you and it showed 2 different people with that same email address and when I went to start a conversation with them it said neither had Google Hangouts. :\textbackslash  My email is [EMAIL] - always on Hangouts

 & Email &   Comments containing user's email.  \\ 
\hline

2) On doit pouvoir le faire de façon un peu plus générique... en demandant si la propriété existe.
Quelque chose du style:
if( getProps().hasProperty( kFnOfxParamPropGroupIsTab )
    getProps().propSetInt( kFnOfxParamPropGroupIsTab, 1 );

 & Language &   Comments in which a user's writes in a language different from English.  \\ 
\hline

3) Is it just me who expected to see the full text from which the “During the ... quite a bit [...]" quote was extracted upon clicking the headline link for this item? Having a direct link to the release download page is wonderful, but can we also have a link somewhere to the “announcement" the text was taken from?

Or is this not an excerpt from anything but an original text?
 & Moral values &   Comments in which there is a moral consideration made by the user that discloses user's ethics or moral values.  \\ 
\hline

4) lib @[USER$_1$] @[USER$_2$] @[USER$_3$] @[USER$_4$] My activity on UMS will be sporadic over the next few weeks, as I have family from overseas visiting. I'll still try to do what I can. Do you think we should release 3.4.0 now?

 & Personal information &   Comments containing any type of personal information not represented by the other labels, for example: information about user's family, trips, studies, habits and so on.  \\ 
\hline

5) This is a good start. Our telephone conversation makes more sense now. I don't think we'll be able to reuse \_exactly\_ same container (`TeamProjectCardContainer`) in the meeting context - that TeamMembers subscription on the container doesn't exactly make sense. We'll probably have to write another container component.

This branch is a good illustration of what we need to do. I didn't notice the `isProject` prop on `OutcomeCard` for example.
& Relation with a user & Comments containing information that reveal a direct relationship with another GitHub user. \\

\hline
6) @[USER$_5$] This is still a very young project. In the latest release we changed the command line syntax as well as I added a `v` to the version number when tagging.  I would have expected more of an outcry from the former rather than passive aggressive comments about the latter. I'm sorry we don't live up to your high standards, but I fail to see how your input here brings the project forward in any constructive manner.  Many other projects don't tag at all, fail to upload tarballs, or if they do said tarballs don't build, and most other projects don't even provide a changelog. Respectfully, comment on things that actually matter.& Community etiquette & Comments containing remarks on how to behave on the GitHub platform.\\

\hline
7) Released. On Fri, Jan 17, 2014 at 12:40 PM, [PERSONAL NAME OF A USER] notifications@github.comwrote:

 Any chance you would release a new version? I just spent two hours tracing
 down this same bug. Thanks.

Reply to this email directly or view it on GitHub [LINK TO A GITHUB PAGE]

[PERSONAL NAME OF A USER]
[COMPANY WEBSITE]

& Workplace & Comments revealing information about user's job or workplace. \\
\hline

8) Basically you can attach release notes to your tags: [LINK TO A GITHUB PAGE]

[PERSONAL NAME OF A USER]

Den 08/02/2015 kl. 22.06 skrev [PERSONAL NAME OF A USER] notifications@github.com:

I haven’t checked out that feature. What’s the scoop?

     On Feb 8, 2015, at 11:54 AM, [PERSONAL NAME OF A USER] notifications@github.com wrote:

    I thought we were switching to the releases feature on GitHub? 
    @[USER$_6$] [LINK TO A GITHUB PAGE]
    @[USER$_7$] [LINK TO A GITHUB PAGE]

    Reply to this email directly or view it on GitHub [LINK TO A GITHUB PAGE].

Reply to this email directly or view it on GitHub.
& Personal name & Comments showing personal name of the user (not their account name). \\

\hline
9) Yeah, we're good. I'll see if I can get posterior prediction with new
levels working for real.

On Sun, Dec 27, 2015 at 7:51 AM, [PERSONAL NAME OF A USER] notifications@github.com
wrote:

 Hope you and your family avoided the worst of the tornado.
 
 On Sun, Dec 27, 2015 at 12:48 AM, bgoodri notifications@github.com
 wrote:
 
  @[USER$_8$] [LINK TO A GITHUB PAGE]
  I pushed a new\_thing() function,
  which is not complete but works like lme4's predict.merMod(). It sort of
  works but needs to be properly integrated into posterior\_predict. I would
  finish it off, but I am about to be hit by a tornado. Reply to this email directly or view it on GitHub [LINK TO A GITHUB PAGE]. & Location & Comments reporting any information that can be linked with the user's location. \\

\hline

\end{tabular}

}

\label{tab: CommentsLabels}

\end{table}
\FloatBarrier

%% file: tex/prediction.tex
\revised{
Pre-trained generic language models~\citep{devlin2018bert,inan2023llama,howard2018universal} have
achieved great results on different NLP tasks.
To illustrate the potential of these models in enhancing user privacy awareness, our study concentrates on demonstrating their ability to detect possible privacy leaks within textual comments.
The primary aim is not to develop high-performance tools or methodologies but rather to explore the feasibility of autonomously detecting self-disclosure across various contexts.} 

\revised{To achieve this, we fine-tuned the Llama2 model~\citep{touvron2023llama} (Section~\ref{sec:llama}), 
using the labeled corpus of comments obtained through the process explained in Sections~\ref{subsec:corpus} and \ref{subsec:manual_labeling}. This fine-tuning process aimed to enhance the model's performance in identifying privacy data leakage, leveraging the targeted information in the curated dataset.}

\revised{Additionally, we investigated the performance of BERT (Bidirectional Encoder Representations from Transformers)~\citep{devlin2018bert} for sensitivity detection (Section~\ref{sec:bert}). BERT, a widely used pretrained model, was fine-tuned for binary classification of self-disclosure in comments. We utilized the Hugging Face Transformers library and trained BERT on our labeled dataset curated in Section~\ref{subsec:manual_labeling} to adapt it to the sensitivity detection task. The task consists of a binary classification of sensitive or insensitive comments.}

\subsection{Large language models for sensitivity detection}\label{sec:llama}

Research in the domain of content sensitivity detection and privacy leakage in text spans various disciplines, including machine learning (ML), natural language processing (NLP), philosophy, psychology, and the social sciences. The overarching goal is to enhance privacy awareness and conduct risk assessments, primarily within specific platforms, with the ultimate aim of developing technology for empowering users in protecting their privacy. 
Despite the considerable success of these approaches, they generally do not conform to a one-size-fits-all model \citep{nguyenFineTuningLlamaLarge2023a,tangDoesSyntheticData2023}. Performance disparities exist across datasets, with models excelling in specific contexts while underperforming in others, as demonstrated in experiments by \cite{peirettiDetectionPrivacyHarmingSocial2023}. Language, dataset balance, and text length are among the factors influencing model effectiveness. Moreover, achieving satisfactory performance necessitates a harmonious combination of feature extraction and classifier design. Researchers are required to explore numerous combinations to optimize synchronization \citep{nguyenFineTuningLlamaLarge2023a,tangDoesSyntheticData2023}. This comprehensive approach is crucial for content sensitivity detection in text.
Recent advancements in LLMs have significantly enhanced the performance of diverse natural language processing (NLP) tasks \citep{10.1145/3605943}, opening up new possibilities for automating functions traditionally executed by humans. These models consist of large neural networks pretrained on vast corpora of text data in multiple languages, offering potential solutions to the challenges associated with conventional text classification methods. This is the reason why we chose to use LLMs to for this detection sensitivity task. \revised{In order to demonstrate the potential of Llama, we have asked the model to provide only Yes or No as an output. In particular, Listing \ref{lst:zero-shot} shows this instruction.
}
\begin{lstlisting}[language=Python, caption={Zero-shot Llama model.}, label=lst:zero-shot,captionpos=t]
def query_llama(comment: str) -> str:
    return f"""### Instruction: We need to classify the text as privacy sensitive or not, and please use Yes or No.

### Input:
{comment.strip()}""".strip()
\end{lstlisting}
Several successful research endeavors have utilized fine-tuning of large language models (LLMs)~\citep{10299938,10031034,YAO2024100211}. In order to enhance the accuracy of Llama's predictions, we have chosen to utilize the Parameter-Efficient Fine-Tuning (PEFT) method. It allows for the effective customization of pre-trained language models (PLMs) for different downstream applications without the need to fine-tune all of the model's parameters. Optimizing extensive pre-trained language models (PLMs) is frequently too expensive. PEFT approaches specifically focus on fine-tuning a limited number of additional model parameters, resulting in a significant reduction in both computational and storage expenses. 
The fine-tuning process involves providing a cue to the model and directing it to provide an appropriate binary classification, specifically distinguishing between privacy-sensitive and privacy-non-sensitive. The target provided corresponded to the projected classification. This was done to enable the model to provide a direct response using binary classification. The prompt employed during the process of fine-tuning is illustrated in Listing~\ref{lst:fine-tune}. It consists of three main parts: \emph{(i)} \textit{Instruction}, where we are asking for a binary classification; \emph{(ii)} \textit{Input} is the comment to be classified; and \emph{(iii)} \textit{Response} is the expected answer.
The results of this experiment are illustrated in Section \ref{sec:EmpiricalResults}.
\begin{lstlisting}[language=Python, caption={Fine-tune the Llama model with the labeled corpus.}, label=lst:fine-tune,captionpos=t]
def generate_training_prompt(comment: str, bin_class: str) -> str:
    return f"""### Instruction: We need to classify the text as privacy sensitive or not, and please use Yes or No.

### Input:
{comment.strip()}

### Response:
{bin_class}
""".strip()
\end{lstlisting}

\subsection{BERT for sensitivity detection}\label{sec:bert}

In our study, we explore the capability of BERT to classify user privacy disclosures effectively. 
Its architecture is based on a transformer encoder and the basic BERT architecture consists of different attention-based layers. The tokens are represented as input vectors which include the tokens they-self, their positions, and their context sentence. For each token, the attention-based layers produce a representation. Each token representation is based on the representations of all tokens.
The output of one attention-based layer is provided as input for the next one. Finally, the last attention layer provides the model output.
The BERT model is unsupervisedly trained on large amounts of text and may later be applied to potentially any task. This allows us to train the \code{bert-base-uncased} model\footnote{\url{https://huggingface.co/google-bert/bert-base-uncased}} using the labeled set of comments defined in Section~\ref{subsec:manual_labeling}. 
We have chosen this model because previous authors have successfully employed BERT in the context of empowering users \citep{adhikari2022privacy,khalajzadeh2022diverse,wang2022personalizing}. The applications span from automated labeling of GitHub issues to privacy policy classification.

\subsection{Metrics and methodology}\label{subsec:metrics}
\revised{Our study evaluates the effectiveness of pre-trained models in identifying private information within text by conducting three distinct configurations.}

\revised{The first configuration, zero-shot Llama (\llamazs), employs the LLaMA model utilizing predefined queries as outlined in Listing~\ref{lst:zero-shot}. This approach tests the baseline ability of LLaMA to recognize privacy-related information without any model customization.
In the second configuration, fine-tuning Llama (\llamaft), we enhance the LLaMA model's capability by fine-tuning it on 80\% of our curated dataset as outlined in Listing~\ref{lst:fine-tune}. The remaining 20\% of the data serves as a test set to evaluate the model’s prediction accuracy. 
The third configuration, fine-tuning BERT (\bertft), involves a BERT model that we trained specifically for the task. Similar to the \llamaft configuration, we fine-tuned the model on 80\% of the manually curated comments, reserving the remaining 20\% for performance evaluation.}

To asses the performance of the model, we used state-of-the art metrics: accuracy, precision, recall, and false positive rate (FPR). In what follows, $TP$ is the number of true positive predictions, $TN$ is the number of true negative predictions, $FP$ is the number of false positive predictions, and $FN$ is the number of false negative predictions. The metrics are calculated as follows \citep{hossin2015review}:

\paragraph{Accuracy} refers to the ratio of right predictions, including both true positives (TP) and true negatives (TN), to the total number of cases analyzed:
\begin{equation}\label{accuracy}
  \text{Accuracy} = \frac{\text{TP} + \text{TN}}{\text{TP+TN+FP+FN}}
\end{equation}

\paragraph{Precision} measures the fraction of the number of correctly classified comments as yes (TP) to the total number of classified comments as yes (TP + FP):
\begin{equation}\label{precision}
  \text{Precision} = \frac{\text{TP} }{\text{TP} + \text{FP}}
\end{equation}

\paragraph{Recall} measures the proportion of actual positive cases that are correctly identified by the model. It is defined as the ratio of true positives (TP) to the sum of true positives and false negatives (FN):
\begin{equation}\label{recall}
\text{Recall} = \frac{\text{TP}}{\text{TP} + \text{FN}}
\end{equation}

\paragraph{F1} represents the harmonic mean between recall and precision values, and it is particularly high when true negatives ($TN$) are high:

\begin{equation}\label{F1}
F1 = 2 \times \frac{\text{Precision} \times \text{Recall}}{\text{Precision} + \text{Recall}}
\end{equation}

Section~\ref{subsec:rq4} shows the results of the different experiments and compare the performances of each configurations.

%% file: tex/empirical_results.tex
In this section, we report and analyze the experimental results by answering the four research questions introduced in Section \ref{sec:Introduction}.



\subsection{\rqfirst}\label{subsec:RQ1}

\begin{figure}[ht]
    \begin{subfigure}{0.4\textwidth}
        \centering
        \includegraphics[width=\textwidth]{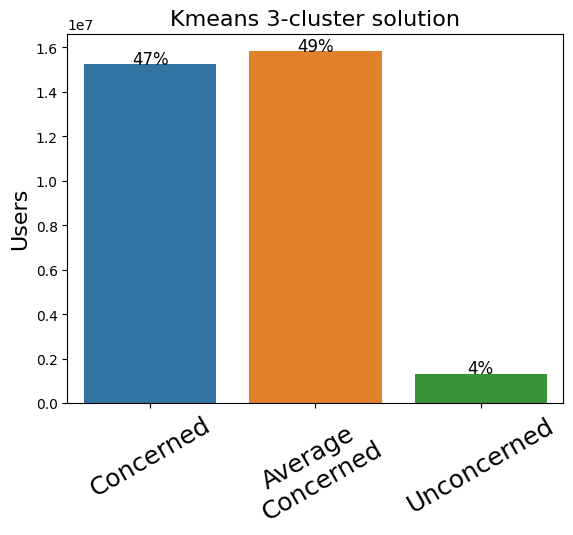}
        \caption{Users per cluster}
        \label{fig:users_SNS3}
    \end{subfigure}
    \hfill
    \begin{subfigure}{0.4\textwidth}
        \centering
        \includegraphics[width=\textwidth]{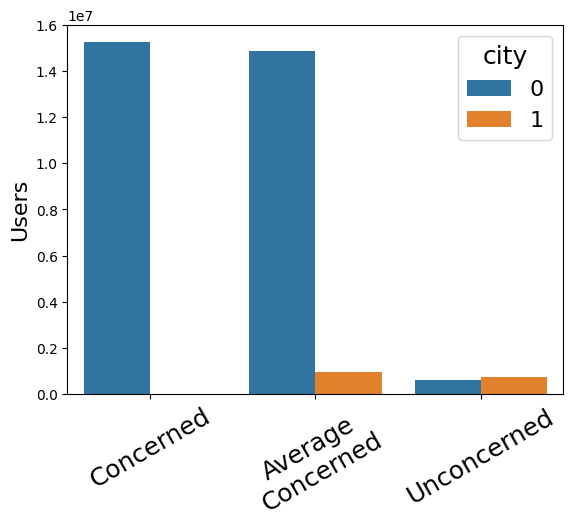}
        \caption{City}
        \label{fig:users_city}
    \end{subfigure}
    \hfill
    
    \begin{subfigure}{0.4\textwidth}
        \centering
        \includegraphics[width=\textwidth]{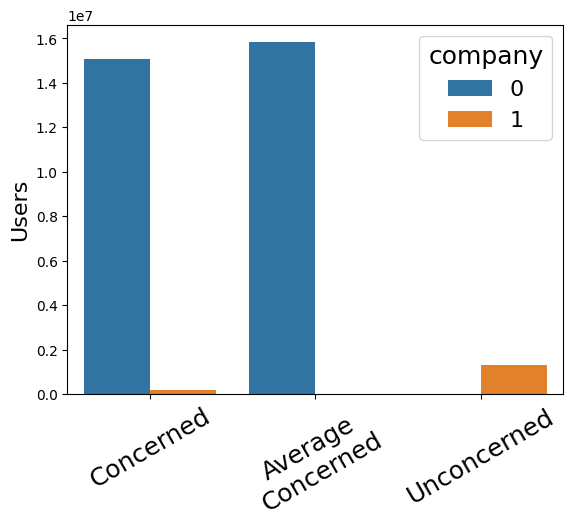}
        \caption{Company}
        \label{fig:users_company}
    \end{subfigure}
    \hfill
    \begin{subfigure}{0.4\textwidth}
        \centering
        \includegraphics[width=\textwidth]{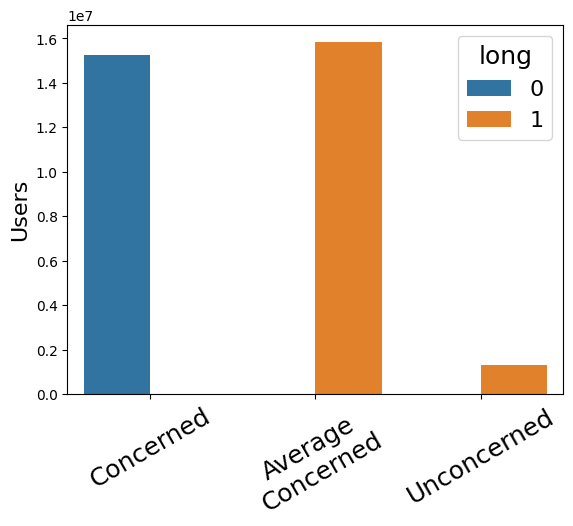}
        \caption{Longitude}
        \label{fig:users_long}
    \end{subfigure}
    
    \caption{\revised{Cluster analysis on the \textit{Users'} dataset with K-means. }
    Number of users per cluster (Figure \ref{fig:users_SNS3}) and distribution of variables per cluster (Figures (\ref{fig:users_city},\ref{fig:users_company} and \ref{fig:users_long}).}
    \label{fig:privacy profiles}
\end{figure}

\paragraph{Users' dataset cluster analysis} 
The cluster analysis conducted on the \textit{Users'} dataset led to three unbalanced clusters, shown in Figure \ref{fig:users_SNS3}. \revised{Figure \ref{fig:users_city}, \ref{fig:users_company} and \ref{fig:users_long} show the number of users that set (1) or hide (0) the corresponding setting being \texttt{City}, \texttt{Company}, and \texttt{Longitude}}. From these plots, 
we can observe that the clusters are distinguished according to each variable considered. For example, \revised{users in the first cluster, named ``Concerned'', exhibit a reluctance to share any information on \GH. Conversely, those in the second cluster, denoted as ``Average Concerned'' are willing to share information only regarding their location (\texttt{Longitude}). On the other hand, users in the third cluster, namely ``Unconcerned'', have a tendency to share comprehensive information through their privacy settings. Indeed, the majority of the users in this cluster have set the option to disclose both their \texttt{City}, their \texttt{Company}, and their \texttt{Longitude} (location).} 
This solution suggests that the privacy desiderata of GitHub users can vary even on a small set of privacy options and it demonstrates that \GH users actively utilize privacy settings. \revised{It is interesting to observe the population distribution depicted in Figure \ref{fig:users_SNS3}, as the cluster ``Unconcerned'' stands out as the less populated cluster \revised{(4\% of the entire population)}. This observation potentially implies that only a limited percentage of \GH users are willing to disclose comprehensive information in their profiles.} This finding further motivates our study.

\begin{figure}[!h]
\begin{subfigure}{0.3\textwidth}
        \centering
        \includegraphics[width=\textwidth]{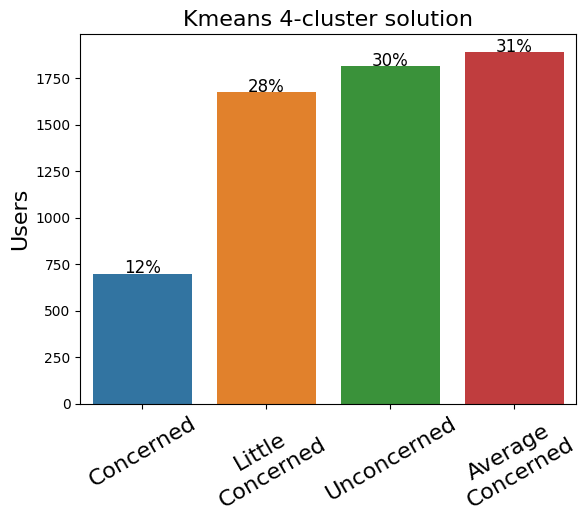}
        \caption{Active users per cluster}
        \label{fig:SNS4}
    \end{subfigure}
    \hfill
    \begin{subfigure}{0.3\textwidth}
        \centering
        \includegraphics[width=\textwidth]{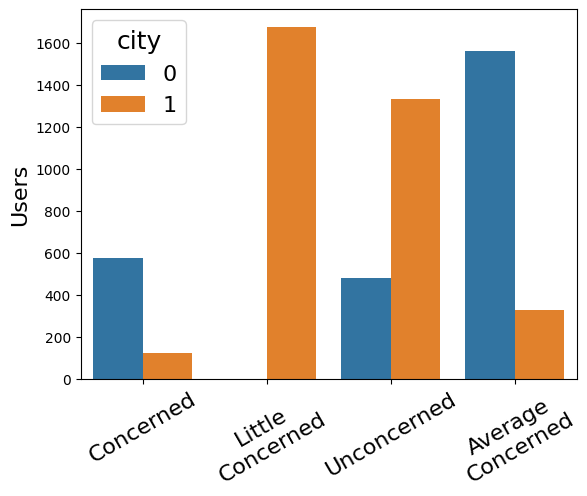}
        \caption{City}
        \label{fig:city}
    \end{subfigure}
    \hfill
    \begin{subfigure}{0.3\textwidth}
        \centering
        \includegraphics[width=\textwidth]{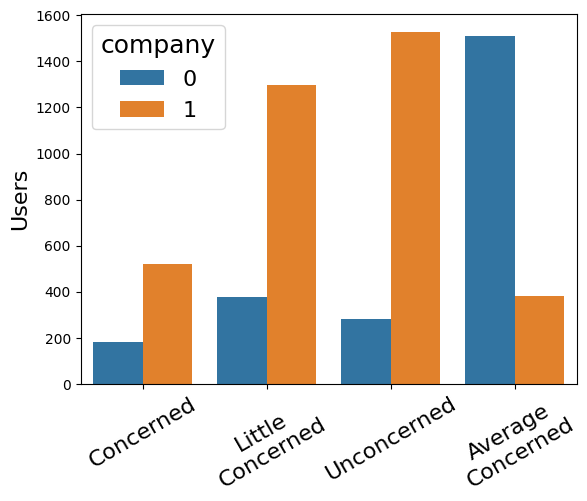}
        \caption{Company}
        \label{fig:company}
    \end{subfigure}
    \hfill
    \begin{subfigure}{0.3\textwidth}
        \centering
        \includegraphics[width=\textwidth]{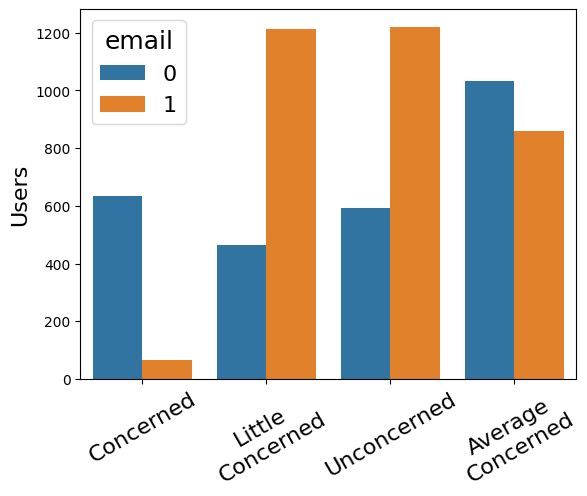}
        \caption{Email}
        \label{fig:email}
    \end{subfigure}
    \hfill
    \begin{subfigure}{0.3\textwidth}
        \centering
        \includegraphics[width=\textwidth]{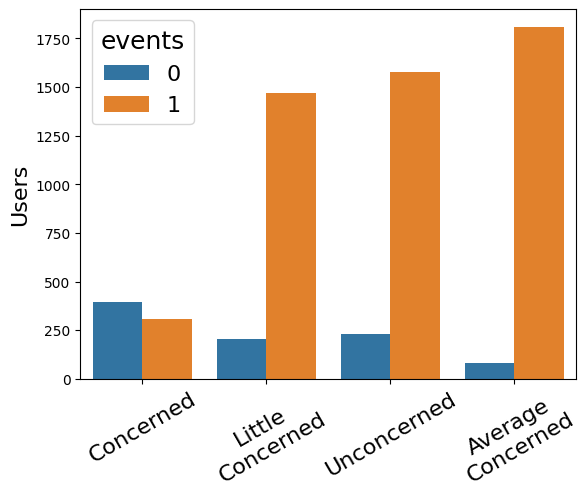}
        \caption{Events}
        \label{fig:events}
    \end{subfigure}
    \hfill
    \begin{subfigure}{0.3\textwidth}
        \centering
        \includegraphics[width=\textwidth]{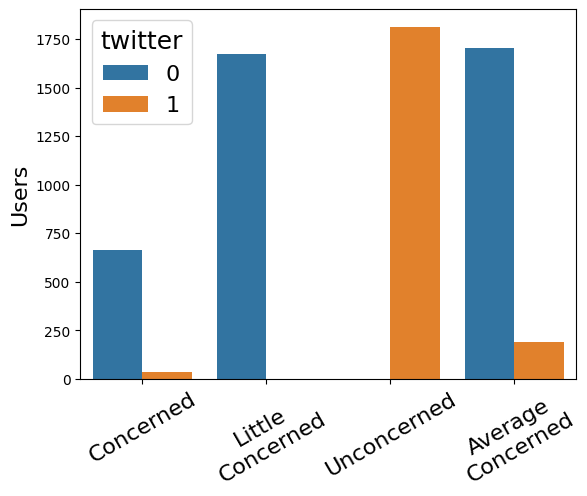}
        \caption{Twitter}
        \label{fig:twitter}
    \end{subfigure}
    \hfill
    
    \caption{\revised{Cluster analysis on the \textit{Active users} dataset with K-means.} Number of active users per cluster (Figure \ref{fig:SNS4}) and distribution of variables per
cluster (Figures \ref{fig:city}, \ref{fig:company}, \ref{fig:email}, \ref{fig:events} and \ref{fig:twitter}).}
   \label{fig:overview of active}
   \vspace{-3pt}
\end{figure}

\paragraph{Active users analysis with K-means} 
On the \textit{Active users} dataset, we performed a K-means cluster analysis with K=4 chosen with the Elbow method. Figure \ref{fig:overview of active} shows an overview of privacy settings choices made by the active users. The privacy profiles are rather balanced, as illustrated in Figure \ref{fig:SNS4} by the cardinality of each cluster. 
Figures \ref{fig:city}, \ref{fig:company}, \ref{fig:email}, \ref{fig:events}, and \ref{fig:twitter} depict users' privacy settings choices per cluster. \revised{We used this analysis to qualitatively define each profile as follows: ``Concerned” \revised{about hiding their information from the GitHub profile}, ``Little concerned'', ``Unconcerned'' and ``Average concerned''. }For instance, Figure \ref{fig:city} reveals that in \revised{the first cluster, namely ``Concerned"}, a significant majority of users opted to conceal their \texttt{City} information on their profiles, \revised{similarly to users in cluster ``Average concerned''. On the contrary, users from clusters ``Little concerned'' and ``Unconcerned'' exhibit a significant number of users willing to share information about their City. Analogous considerations can be applied to the other variables: \texttt{Company}, \texttt{Email}, \texttt{Events} and \texttt{Twitter}.}

These results are significant as they illustrate that users' privacy concerns, as expressed through the privacy settings of \GH, vary. Thus, users exhibit distinct privacy profiles. 
This is visible in Figure \ref{fig:SNS4}, where the profiles are rather balanced in terms of the number of users, showing that the choices of privacy settings do not converge towards one main combination of settings. 
These privacy profiles helped us categorize active users according to their privacy desiderata. We exploited this categorization 
to address RQ$_3$. 

\begin{figure}[ht]

\begin{subfigure}{0.3\textwidth}
        \centering
        \includegraphics[width=\textwidth]{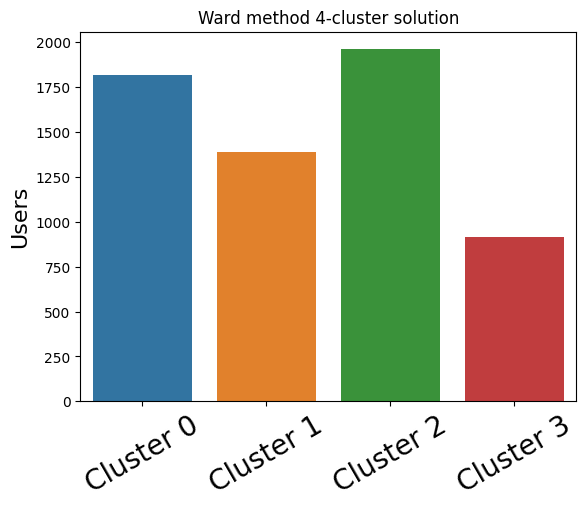}
        \caption{Active users per cluster}
        \label{fig:ward}
    \end{subfigure}
    \hfill
    \begin{subfigure}{0.3\textwidth}
        \centering
        \includegraphics[width=\textwidth]{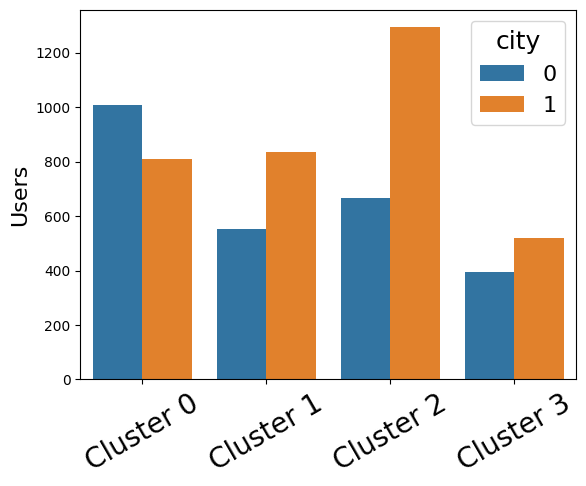}
        \caption{City}
        \label{fig:wardcity}
    \end{subfigure}
    \hfill
    \begin{subfigure}{0.3\textwidth}
        \centering
        \includegraphics[width=\textwidth]{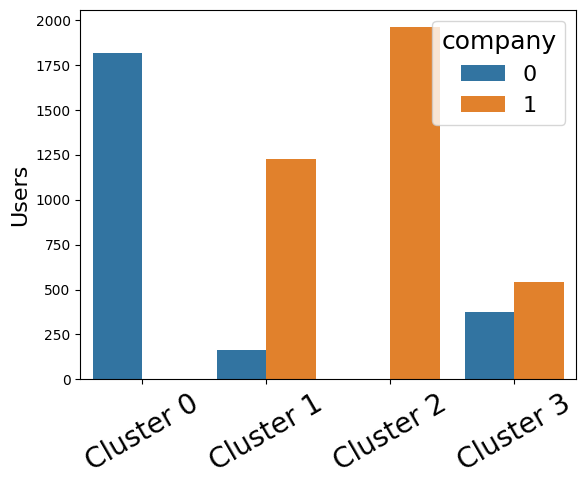}
        \caption{Company}
        \label{fig:wardcompany}
    \end{subfigure}
    \hfill
    \begin{subfigure}{0.3\textwidth}
        \centering
        \includegraphics[width=\textwidth]{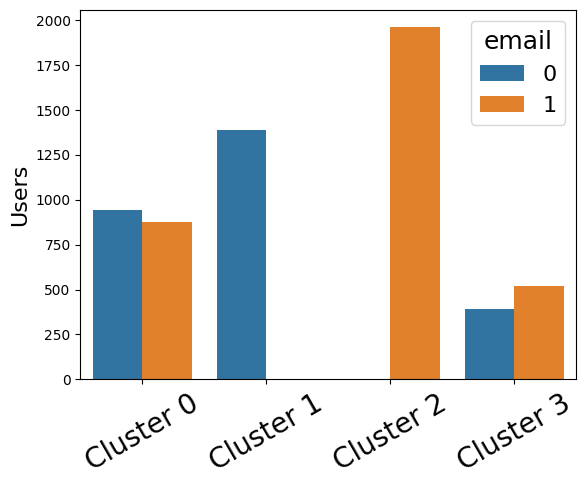}
        \caption{Email}
        \label{fig:wardemail}
    \end{subfigure}
    \hfill
    \begin{subfigure}{0.3\textwidth}
        \centering
        \includegraphics[width=\textwidth]{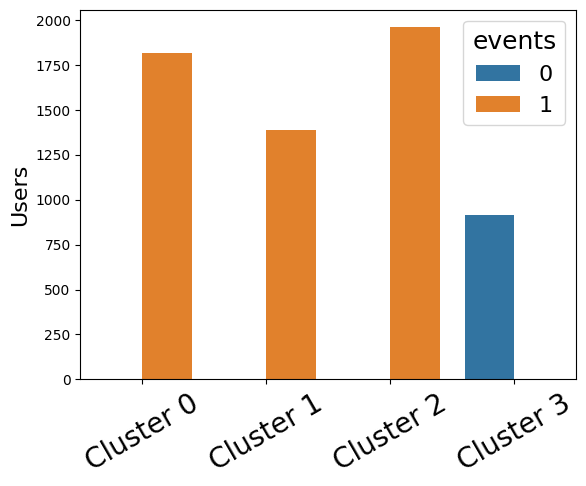}
        \caption{Events}
        \label{fig:wardevents}
    \end{subfigure}
    \hfill
    \begin{subfigure}{0.3\textwidth}
        \centering
        \includegraphics[width=\textwidth]{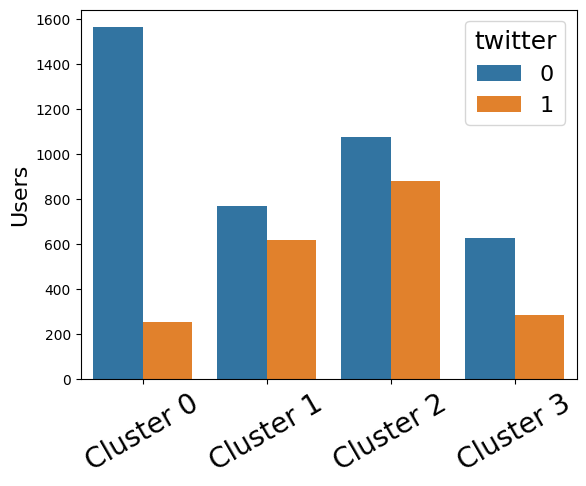}
        \caption{Twitter}
        \label{fig:wardtwitter}
    \end{subfigure}
    \hfill
    
    \caption{\revised{Cluster analysis on the \textit{Active users} dataset through the hierarchical clustering methods.} Number of active users per cluster (Figure \ref{fig:ward}) and distribution of variables per
cluster (Figures \ref{fig:wardcity}, \ref{fig:wardcompany}, \ref{fig:wardemail}, \ref{fig:wardevents} and \ref{fig:wardtwitter}).}

   \label{fig:overview_active_ward}
   \vspace{-10pt}
\end{figure}

\paragraph{Active users analysis with hierarchical clustering} 
As many authors suggest \citep{brandao2022prediction, sanchez2020recommendation}, another way to generate users' profiles is through a hierarchical clustering algorithm. We used this method to further analyze the \textit{Active users} dataset. We applied this technique using Ward's method on the \textit{Active users} dataset, with the number of clusters equal to 4. As previously explained, this number was chosen by analyzing the dendrogram (Figure \ref{fig:ward_dendro}). We report the bar charts regarding the distribution of variables per cluster and the cardinality of each cluster (see Figure \ref{fig:overview_active_ward}). The clusters are less balanced than the one obtained with K-means (Figure \ref{fig:ward}). By observing the distribution of variables per cluster, the situation is unclear compared to the clusters with K-means. Indeed, the variable \texttt{City} seems irrelevant in discriminating between the profiles (Figure \ref{fig:wardcity}).

\begin{quote}
\textbf{Answer to RQ$_1$.} GitHub users manifest diverse privacy preferences, as reflected in their selection of privacy settings, both on a broad scale —exemplified by the analysis conducted on the entire \textit{Users} dataset— and on a more granular level, as seen in the examination of the \textit{Active user} dataset. The latter is interesting, given that these users share a high level of activity on the platform, yet their privacy preferences can vary considerably. According to our analysis of the \textit{Users} dataset, it emerges that only a small percentage of users are willing to disclose all their information in their \GH profiles. \revised{Overall, privacy settings are a tool used by users to safeguard their privacy and should accurately reflect their privacy preferences.}
\end{quote}

\subsection{\rqsecond}

\revised{In order to address RQ$_2$, we began with a dataset of 2,000 texts of \textit{Active users}, selected using the Privacy Dictionary, as described in Section \ref{subsec:corpus}. From this corpus, we manually labeled 147 comments as privacy sensitive.}

\begin{figure}
    \centering
    \includegraphics[scale=0.68]{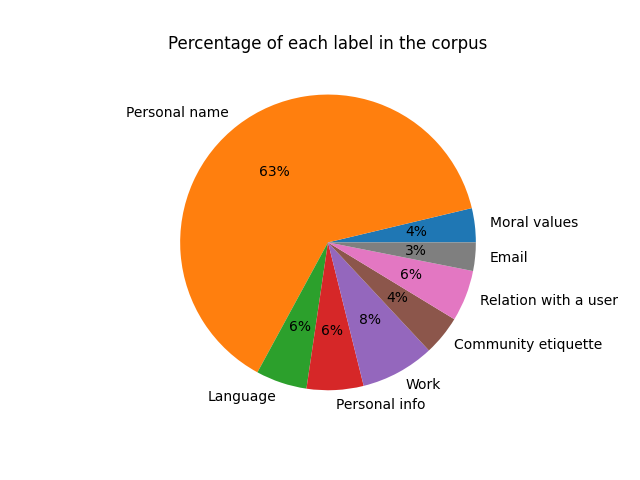}
    \caption{\revised{Percentage of each label in the corpus.}}
    \label{fig:piechart}
\end{figure}

This corpus provides examples of different types of private information that is disclosed by the \GH users. Sometimes this information is more explicit, as visible in comment 1) from Table \ref{tab: CommentsLabels}. In this case, the user reveals his/her personal email. Sometimes, a piece of private information is more implicit, as in comment 8), where the user speaks of a tornado in the area of his/her interlocutor. This reveals a close relationship between the users and the information of the tornado can lead to the location of the interlocutor. Similarly, comment 5) reveals a close relationship between the two users, having a conversation on the phone. 
\revised{Figure \ref{fig:piechart} shows the percentage distribution of each label in the corpus. As it is evident, ``Personal name" represents a significant portion of the pie chart. However, it is noteworthy that nearly 40\% of the sensitive comments contain a wide range of sensitive information, from ``Moral values" to work-related details. }

This finding reaffirms the observations made by previous researchers, that while GitHub primarily serves as a platform for sharing technical knowledge, an examination of users' textual comments exposes instances of disclosing private information. Such information may originate from the commenter, either pertaining to themselves or involving another user. 
\begin{quote}
\textbf{Answer to RQ$_2$.} Even if \GH is considered a platform used for technical purposes only, different types of private information are consciously or unconsciously disclosed by the users. 
The categories of information revealed in pull\_request comments exhibit diversity. To date, our investigations have uncovered instances of \textbf{Personal Name}, \textbf{Workplace}, \textbf{Email}, \textbf{Location}, \textbf{Moral Values}, \textbf{Community Etiquette}, \textbf{Personal Info}, \textbf{Language}, and \textbf{Relation with a User}.
\end{quote}

\subsection{\rqthird}

During the manual-labeling process described in Section \ref{subsec:manual_labeling}, we selected comments that were particularly meaningful from a privacy perspective to analyze the profile of the author. \revised{Table \ref{tab: comments and profile} presents examples of sensitive comments and the profiles of their authors, as identified in Section \ref{subsec:RQ1}. }

\FloatBarrier
\begin{table}[ht]
\caption{Examples of comments in each profile.}
\adjustbox{max width=\textwidth}
{
\begin{tabular}{p{11 cm}p{3.7 cm}}
\hline
\textbf{Comments}  & \textbf{Privacy Profile}   \\
\hline
My two cents. 

This sure would be a very different conversation if it was 'WebM for Niggers' or 'WebM for Kykes'. Those words are more accepted as being derogatory in society. I think a lot of people use 'retard' as a colloquial, playful word, but really, I think it isn't acceptable; a community, however minority, is [voicing it's opinion]([LINK TO A WEBSITE]]) more openly about this. I think they are right.

We must, I believe, be wary and think critically about any form of censorship. In this case, I believe the censorship is legitimate because the change does not affect any functionality of the repository or suppress opinions (blowing up the forks was quite silly though). Each case of censorship should be exposed to this amount exposure and community thought. I think we are doing well in that regard.

Github are enforcing their TOS and I believe this change is for the positive. People implying that Github are restricting their freedom, I think, will need to try and empathise more with the community of people affected by this word.
 &   CONCERNED  \\ 
\hline
@[USER$_7$] Hey Gil, I am sorry if my comment sounded harsh to you. I know you are working your ass off for the community and I really appreciate your work here. Applying changes to all our packages is a boring and unthankful job, I can tell that from own experience. I understand that you don't have time to create pull requests in that situation and I am sorry that I did not take this into account when asking for a PR.

I got frustrated lately with plone.restAPI development a bit because other devs with good intentions do things that lead to additional work for me as a maintainer.

plone.restAPI uses semantic versioning and my goal is to do very frequent releases after every single meaningful pull request. This is only possible if I can merge PRs at any time and then do a release right away. This wasn't possible because of your commit to master. With a PR I could have easily postponed the merge of this important and valuable contribution to after the next release.

The simple reason why I did not complain about Maurits PR is that this one did not require me to manually go through all open PRs and having to amend them.
   &  LITTLE CONCERNED  \\

\hline
A lot of thought and effort has gone into researching the subject of ”deep overriding” by Russell O'Connor in the `haskellPackagesFixpoint` branch and by myself in the `haskell-ng` branch. You are aware of these activities, but apparently chose to ignore them and instead committed your own ”deep overriding” solution to `master` without any prior consultation. There is nothing inherently ”wrong” with doing that; you are entitled to commit whatever changes you feel are best. Personally, I perceive this kind of unilateral decision making as disrespectful towards other stakeholders, though. Also, I feel that your solution is weird from a technical point of view, and having it reviewed by others before you committed certainly wouldn't have hurt the quality/readability/re-usability of your patch.
 &    UNCONCERNED  \\ 

\hline
Hi [PERSONAL NAME OF A USER],

Of cause I follow your development. I have a script running which checks
genode.org for changes. The release notes are like early Christmas presents :)
I would like to go to FOSDEM next year, but I'm not sure if Daniel will 
be able to come with. Are you attending? It would be amazing to meet 
up with you guys again.

Cheers
/u

 &  AVERAGE CONCERNED 
   \\ 
\cline{1-1}
To be honest, I just missed it when I was reviewing the release notes @[USER$_8$] had created in the first place. I don't mean to be political hence the removal since whether we like it or not that phrase brings certain things to mind. & \\
\hline              
\end{tabular}
}
\label{tab: comments and profile}
\end{table}
\FloatBarrier

\begin{figure} 
    \centering 
    \begin{subfigure}[b]{0.9\textwidth} 
        \includegraphics[scale=0.6]{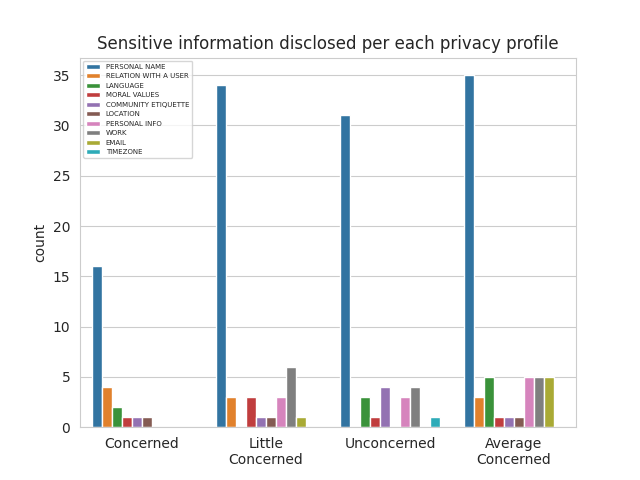}
        \caption{Distribution of label in each privacy profile.}
        \label{fig:overview_sens}
    \end{subfigure}
    \hspace{0.05\textwidth} 
    
    \begin{subfigure}[b]{0.9\textwidth} 
        \includegraphics[scale=0.6]{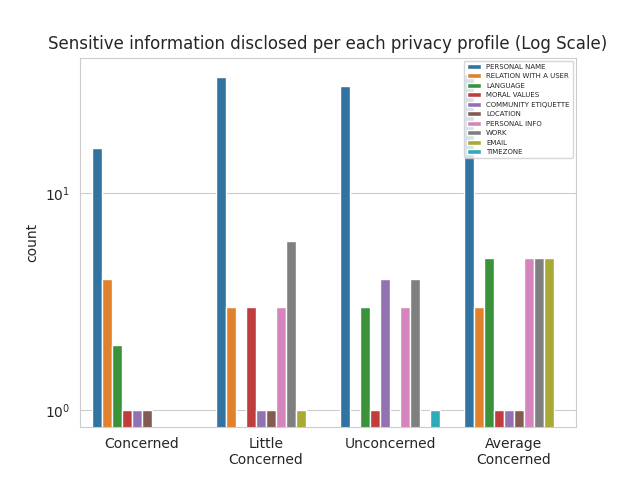}
        \caption{Distribution of label in each privacy profile with the y-axis logarithmically scaled.}
        \label{fig:overview_sens_log}
    \end{subfigure}
    
    \caption{Analysis of the disclosure of each label per privacy profile.}
    \label{fig:overview_sens_combined}
\end{figure}

Interestingly, many of these comments fall into the profiles of ``Average Concerned'' and ``Concerned'' users. \revised{This suggests that users who are presumed to be concerned about their privacy do not necessarily demonstrate this concern through their behaviors.} \revised{Figure \ref{fig:overview_sens_combined} shows the distribution of each label per privacy profile, i.e., how many comments disclosing that information were found in each cluster. The bar plot in Figure \ref{fig:overview_sens} represents the distribution of sensitive comments across different privacy profiles. Each privacy profile is indicated on the x-axis, categorized into the four groups: ``Concerned", ``Little Concerned", ``Unconcerned", and ``Average Concerned". The bar plot in Figure \ref{fig:overview_sens_log} presents the same data with the y-axis logarithmically scaled, allowing for a more compact and interpretable visualization. 
Consistent with previous findings, the label ``Personal name" is the most commonly shared across all privacy profiles. As shown in the plot, the profile of ``Concerned" users displays six out of nine labels, which is expected as these users are likely more attentive to their privacy. However, evidence of self-disclosure is still found among users in this cluster. Conversely, the ``Average Concerned" user profile contains all the different labels, suggesting a discrepancy between their stated privacy preferences and their actual behaviors.} 

This phenomenon can be interpreted in various ways. One possible explanation is that \GH privacy settings may not comprehensively capture users' privacy preferences. Alternatively, developers might believe that disclosing certain information could be advantageous in specific situations, potentially overlooking the fact that this information is openly available. Another consideration is the so-called ``privacy paradox" \citep{acquistiPrivacyRationalityIndividual2005,gerberExplainingPrivacyParadox2018}. For example, users in the profile ``Average Concerned" express a desire to keep information private through their privacy choices but exhibit a disclosing behavior in their writing/commenting activities. However, the validity of the privacy paradox is highly debated in the literature, and its existence is questioned \citep{solove2021myth}. Lastly, this finding can be also explained with the concept of ``privacy fatigue", which refers to the increasing difficulty individuals face in managing their online personal data, leading to weariness about having to constantly consider online privacy \citep{choi2018role}.

\begin{quote}

\textbf{Answer to RQ$_3$.} The privacy preferences selected by the users on GitHub might not be entirely representative of their privacy desiderata. Indeed, we found a discrepancy between what they declared as privacy settings, and how they behaved. \revised{This finding can be attributed to various factors, spanning from a perceived advantage in sharing specific information, to the so-called ``privacy fatigue", to a potential lack of awareness, often referred to as the ``privacy paradox".}
\end{quote}


\subsection{\rqfourth}\label{subsec:rq4}
\revised{This section compares the configurations described in Section~\ref{subsec:metrics}, each fine-tuned or used directly to classify privacy disclosures in textual comments. Our analysis focuses on several key performance metrics and explores the models' practical effectiveness in real-world privacy identification tasks.}

In Table \ref{tab:prediction_score}, we present the results of configurations in which comments were classified as either privacy-sensitive (PS) or non-sensitive (PNS). 
\revised{In particular, the table compares the three configurations described in Section~\ref{sec:prediction}, \ie \llamazs, \llamaft,  \bertft, across four metrics described in Section \ref{subsec:metrics}.
Evidently, by the \llamaft and \bertft configurations, the prediction performance is better than that of \llamazs. The accuracy metrics stand at 0.94, 0.94 and 0.417, respectively, signifying that the fine-tuned configurations outperform the zero-shot in predicting the correct class. Across all the configuration, it is noticeable that the values for predicting non-sensitive instances consistently outweigh those for sensitive ones. This imbalance can be attributed to the fact that the dataset used is heavily skewed towards non-sensitive texts. Further improvement can be achieved with a more balanced dataset. }

\revised{Even though we have asked the model to provide only ``Yes" or ``No" as acceptable answers (see Listing~\ref{lst:zero-shot}), we observed that the output of the \llamazs was frequently uninterpretable. Examples such as ``I'm not sure what you mean by 'privacy-sensitive,' I'm not sure what you mean by 'not'" and ``This is a simple yes or no" were generated by the model. On the contrary, the \llamaft always generated interpretable output, with a clear answer to the prompt given which consisted in either ``Yes" or ``No". Further studies should delve into prompt-engineering techniques to assess whether the performance in predicting sensitive text can be significantly enhanced. 
It is worth noting that by \llamaft and \bertft their prediction performance is somehow comparable, \revisedtwo{indicating that both models can be exploited for building a tool of sensitivity detection. The curated dataset together with the python scripts to preprocess and fine-tuning pre-trained models are available on the supporting GitHub repositories.\footnote{\url{https://github.com/MDEGroup/EMSE-CHASE-Privacy}}}}


\begin{quote}
    \revised{\textbf{Answer to RQ$_4$.} Pre-trained models can be employed to identify sensitive information in textual data. The availability of a curated corpus of user comments enables the fine-tuning of pre-trained models. Both \llamaft and \bertft configurations exhibit superior performance compared to the \llamazs. Nevertheless, to ascertain whether the low performance on sensitive data is attributed to the skewed dataset or the model itself, it is recommended that more examples of sensitive texts be introduced for evaluation.}
\end{quote}

\begin{table}[t]

    \caption{\revised{Prediction performances for binary classification.}}\label{tab:prediction_score}
    \centering
    \begin{threeparttable}
    \begin{tabular}{l|c|c|c|c|c|c|c|c|c|c|} \hline
                         & \multicolumn{3}{c|}{\textbf{\llamazs}}                                           & \multicolumn{3}{c|}{\llamaft}
                         & \multicolumn{3}{c|}{\textbf{\bertft}} & \\ \hline
                         & Prec. & Recall & F1 & Prec. & Recall & F1 & Prec. & Recall & F1  & Support\\ \hline
    PS$^1$    & 0.69  & 0.35       &    0.46       & 0.59     & 0.38  & 0.46 & 0.52&0.42&0.47& 27    \\ 
    PNS$^2$ & 0.79          & 0.42       & 0.55           & 0.96     & 0.98  & 0.97  &0.96&0.97&0.97& 373   \\ \hline
    Acc.             & \multicolumn{3}{c|}{0.42}                                               & \multicolumn{3}{c|}{0.94} & \multicolumn{3}{c|}{0.94} & 400                                                    \\ \hline
    \end{tabular}
    \begin{tablenotes}
    \item[1] PS stands for the privacy sensitive class
    \item[2] PNS stands for the privacy not-sensitive class
  \end{tablenotes}
    \end{threeparttable}
\end{table}

\vspace{.1cm}
\noindent

%% file: tex/discussion.tex



Our empirical analysis addressed \revised{the study of the privacy dynamics on \GH, with particular attention to users' privacy settings and behaviours on the activities related to pull\_requests comments}. Users privacy preferences were deduced from the privacy settings they chose in their profiles, while the analysis of their behaviours was conducted on their textual activity (pull\_requests comments). Primarily, we observed that users from both \textit{Users} and \textit{Active users} dataset exhibit significantly distinct privacy preferences. \revised{This finding indicates that users actually adopt privacy settings and that they express different privacy concerns (RQ1).} Consequently, there is a clear indication for a more thorough investigation into the dynamics of privacy on this platform.

Additionally, despite \GH being primarily used for sharing technical knowledge, there are instances of unintentional or deliberate leakage of users' private information in their textual activity (RQ2). \revised{The disclosed information ranged from real names to personal values, surpassing what could be concealed by the privacy settings provided on \GH.} 
\revised{This is in contrast with} the \GH privacy statement, which asserts that it is sufficient to ``adjust your setting for your email address to be private in your user profile''
(see Figure \ref{fig:privacystatement}). Indeed, along with previous studies \revisedtwo{\citep{vasilescuGenderTenureDiversity2015,terrell2017gender,meliHowBadCan2019,niuCodexLeaksPrivacyLeaks}}, we realized that this statement does not hold true. 
This suggests that privacy settings alone do not guarantee users' privacy and \revised{that a more sophisticated tool for privacy protection and awareness is needed.}

\revised{The analysis of user behaviors (pull\_request comments) has enabled us to identify diverse types of sensitive information disclosed on GitHub and compare them with the privacy preferences expressed by the users (privacy profile).} We observed that users assigned to a privacy-concerned profile were authors of privacy-sensitive comments (RQ3). \revisedtwo{Our findings indicate that although users do engage with privacy settings using various configurations, their behaviors may inadvertently expose certain private information.} This can be attributed to users' lack of awareness, convenience, or privacy fatigue. In any case, more sophisticated privacy settings on \GH would allow users to more accurately reflect their preferences. 
\revised{
Due to the limitations of privacy settings, we explored using Llama2 and BERT to detect sensitive comments on \GH (RQ4). This preliminary study suggests that BERT outperforms Llama2 in this task. Further investigation with more prompt engineering should be conducted to confirm this result.}
 The implementation of finer-grained privacy settings could serve as a foundation for developing a privacy awareness tool on \GH, which could notify users when their behavior, \revised{identified with the help of models like Llama2 or BERT,} deviates from what is specified in their profile. \revisedtwo{In this context, a privacy awareness tool could be useful for alerting users when such deviations occur or for suggesting less sensitive rephrasings of text. }

\subsection{Threats to validity}


\begin{itemize}
    \item \textbf{External validity.} By using the GHTorrent dataset, we were able to get a large sample of privacy settings on \GH, \revised{as well as the number of actions performed by each user.} This allowed us to establish a definition of \textit{Active users} on the platform. It is worth noting that the GHTorrent version we used was from 2019, which means our selection of \textit{Active users} was based on data available up to that time. \revised{So, no new users were added to the original GHTorrent data and the analysis was done on users who were part of the 2019 dump.} To tackle this potential limitation, we updated users' privacy preferences and their comments using the \GH API. 
Moreover, our results mainly concern the population of \textit{Active users} and might not apply to other \GH users. \revised{Future studies could address this limitation by directly collecting data from the platform and verifying whether the results still hold.}

\revisedtwo{We have excluded comments without privacy-related labels from the fine-tuning and evaluation datasets. While this does not seem to compromise the model's ability to distinguish between sensitive and non-sensitive comments, the lack of these comments may limit the generalizability and robustness of the findings. A more diverse dataset could help address this concern.}

\revisedtwo{The use of truncated comments from GHTorrent may have led to the omission of privacy-related terms in the truncated sections. While this does not affect the accuracy of the automated labeling, it may have resulted in an underestimation of the total number of privacy-sensitive comments.}

The textual data comprised only \PRC. We chose this as ``pull request comments are likely to
contain valuable insight into the relationships of developers interacting with one another''  \citep{sajadi2023interpersonal}, thus we expected to find more sensitive information in this type of data. Future work may consider also 
issues and commits.

\item \textbf{Internal validity.} We adopted privacy settings chosen by the users as a declaration of their privacy desiderata. Even if this can be the case for some of them, for others their choice of privacy settings might be arbitrary or not necessarily aligned with their actual preferences. This discrepancy could stem from limitations in the options available on GitHub or from a lack of awareness regarding privacy implications. Moreover, some users might be unemployed and therefore not showing information related to their job, or not have a Twitter account. 
In our study, these cases are considered as hiding information. 


\item \textbf{Construct validity.} 
While there were four annotators in total, each comment was evaluated by only two individuals. People may perceive the same comment in different ways, and this can pose a threat to the corpus validity. To mitigate this bias, we organized the discussion phase, where any discrepancy was discussed and resolved by two involved participants.
The agreement process took place in two different sessions, during which the raters could explain their choices. 

For a more nuanced understanding of user privacy, a method like the Experience Sampling Method could offer valuable insights \citep{zhang2020understanding, zhang2021did}.

\end{itemize}

%% file: tex/conclusion.tex
With the aim of gaining a better understanding of users' privacy on GitHub, we conducted an empirical study on this platform regarding users' privacy preferences and behaviors. Our findings demonstrate that users actively engage with the platform's privacy settings, leading to the identification of four distinct privacy profiles that emerged from a cluster analysis on the privacy settings (RQ1).
\revised{For what concerns privacy behavior, we found that users share different personal information on \GH, including family details, location, and company-related information, among other things (RQ2). This indicates that on \GH there is a wide range of private information that can be associated with a user, beyond technical matters and surpassing information that can be hidden using privacy settings.}

\revised{
The synthesis of these results revealed a discrepancy between the chosen privacy settings and the actual behaviors of users (RQ3). Despite the possible explanations for interpreting this phenomenon, it is evident that privacy settings alone are insufficient to ensure users' privacy. 
The last result underscores the necessity for more nuanced privacy settings on these platforms and the development of automated tools to assist users in consistently managing their actions on the platform. Given the limitations of privacy settings in protecting users' privacy, we explored adopting models like Llama2 and BERT to detect sensitive comments and pave the way for a privacy awareness tool (RQ4). }

Indeed, our work enables the creation of a privacy awareness tool on \GH that may advice users when they are entering sensitive information in their comments. In this direction, our corpus can be exploited for prediction on sensitive comments. At a more general level, this study lays the foundation for a methodology to observe similar privacy vulnerabilities also on other platforms.


In our future research, we endeavor to enhance our understanding of the necessity for automated privacy tools on \GH. This will be achieved through the implementation of a user survey, akin to the one made by \cite{vasilescuPerceptionsDiversityGit2015}. \revised{In this respect, it would be valuable to acquire updated data on privacy choices made by the same users in 2024. This could serve as a future work to explore the evolution of privacy attitudes and its potential impact on the use of privacy settings.} 
Furthermore, our objective is to augment the size of the labeled corpus by leveraging Llama2, as fine-tuned in this study, or BERT. We plan to utilize this corpus to create a multi-label tool capable of predicting specific sensitive information disclosed in comments. \revisedtwo{An expanded corpus should also include issue and commit comments to ensure a more diverse range of textual data is represented in the dataset.} 
\revisedtwo{The various types of sensitive information disclosed on this platform can have varying consequences on an individual's life. In our future research, it is crucial to study the impact that each type of self-disclosed information may have on users' lives. This could also enhance the effectiveness of privacy awareness tools.} 

Ultimately, our overarching goal is to develop a tool that suggests sanitized comments upon identifying sensitive information thus empowering the users in managing their privacy concerns. The authors are actively progressing along this trajectory.

%% file: tex/statement.tex
The authors have no relevant financial or non-financial interests to declare.

%% file: tex/das.tex
The experimental data and the simulation results that support the findings of this study are available in \GH in the following address:
\url{https://github.com/MDEGroup/EMSE-CHASE-Privacy}.